\documentclass[11pt,a4paper]{article}
\usepackage{geometry}
\usepackage{setspace}
\usepackage[scaled=0.9]{helvet}
\usepackage{graphicx}
\usepackage{subfig}
\usepackage{amssymb}
\usepackage{rotating}
\usepackage{arydshln}
\usepackage{multicol}
\usepackage{abstract}
\usepackage{ucs}
\usepackage[utf8x,utf8]{inputenc}
\usepackage[T1]{fontenc}
\usepackage[swedish, english]{babel}
\usepackage{fancyhdr}
\usepackage{slashed}
\usepackage{enumitem}

\begin{document}
\bibliographystyle{acm} 
\pagestyle{fancy}
\cfoot{\thepage}
\renewcommand{\abstractname}{}

\title{\fontfamily{phv}\selectfont{\huge{\bfseries{
Particle-level kinematic fingerprints and the multiplicity of neutral particles from low-energy strong interactions
}}}}
\author{
{\fontfamily{ptm}\selectfont{\large{Federico Colecchia}}}\thanks{Email: federico.colecchia@brunel.ac.uk}\\
{\fontfamily{ptm}\selectfont{\large{{\it Brunel University London, Kingston Lane, Uxbridge UB8 3PH, UK}}}}
}
\date{}
\maketitle
\begin{onecolabstract}
The contamination, or background, from uninteresting low-energy strong interactions is a major issue for data analysis at the Large Hadron Collider. In the light of the challenges associated with the upcoming higher-luminosity scenarios, methods of assigning weights to individual particles have recently started to be used with a view to rescaling the particle four-momentum vectors. We propose a different approach whereby the weights are instead employed to reshape the particle-level kinematic distributions in the data. We use this method to estimate the number of neutral particles originating from low-energy strong interactions in different kinematic regions inside individual collision events. Given the parallel nature of this technique, we anticipate the possibility of using it as part of particle-by-particle event filtering procedures at the reconstruction level at future high-luminosity hadron collider experiments.
\end{onecolabstract}

\begin{multicols}{2}
{\bf Keywords:}
29.85.Fj; High Energy Physics; Particle Physics; Large Hadron Collider; LHC; background discrimination; mixture models; latent variable models; sampling; Gibbs sampler; Markov Chain Monte Carlo; Expectation Maximisation

\section{Introduction\label{intro}}

The subtraction of contamination from low-energy physics processes described by Quantum Chromodynamics (QCD) is a critical task at the Large Hadron Collider (LHC). The impact of such a correction is going to become even more significant in the upcoming scenarios whereby the high-energy parton scattering of interest will be superimposed with a higher number of low-energy interactions associated with collisions between other protons, the so-called pileup events.

Pileup results in the presence of multiple vertices inside collision events, which often makes the study of rare processes particularly challenging. Subtraction techniques are well established, and typically combine tracking information for charged particles with estimates of the energy flow associated with neutral particles\footnote{
Whenever neutral particles are referred to in the text, neutrinos are not considered.
} that originate from low-energy QCD interactions \cite{REVIEW_PILEUP_2014}. However, while the information provided by the tracking detectors can significantly ease the task of assigning charged particles a vertex of origin, that of associating neutral particles with a probability of their originating from low-energy QCD interactions as opposed to the high-energy parton scattering of interest is a much harder task. 

A number of pileup subtraction techniques have been proposed over the years and are part of the core reconstruction pipelines at hadron collider experiments. With a view to achieving improved performance at higher luminosity, techniques that work at the level of individual particles inside collision events have recently been proposed \cite{PUPPI,SoftKiller,berta} and are being evaluated at the LHC. In particular, particle weighting methods have been presented whereby individual particles are assigned a probability for their origin in soft QCD interactions as opposed to the signal hard parton scattering. The weights are typically used either to rescale the particle four-momentum vectors \cite{PUPPI} or in conjunction with multiple interpretations of the data \cite{jet-sampling}.

In this article, we propose a different approach whereby the weights are instead employed to reshape the particle-level kinematic distributions inside individual collision events.

We build on a view of events as mixtures of particles originating from different physics processes, namely a signal hard parton scattering and background low-energy strong interactions\footnote{It is worth noticing that, while the idea of assigning individual particles a single process of origin is per se conceptually flawed in hadron collider environments due to the presence of colour connection, the use of particle weights provides the required flexibility in interpreting the origin of individual particles.}.

Due to the quantum nature of the underlying physics, the kinematic distributions of particles originating from a given process, e.g. from low-energy strong interactions, normally exhibit a certain degree of variability across collisions. In other words, individual events can be associated with distinctive particle-level pileup kinematic patterns, or ``fingerprints'', as discussed in section \ref{fingerprints}.

We report on the use of this technique on simulated data and show that our algorithm produces reasonable estimates of the number of neutral pileup particles in different kinematic regions inside events regardless of whether or not particles originating from the hard parton scattering are present. To our knowledge, this is the first method of estimating how neutral pileup particles are distributed in different kinematic regions inside individual events thereby taking into account the inherent variability across collisions.

We expect this technique to improve further on the resolution of the missing transverse energy\footnote{
Missing transverse energy is the event-level energy imbalance measured on a plane perpendicular to the direction of the colliding particle beams.
}, as well as on the estimates of particle isolation in higher-luminosity regimes.

Missing transverse energy plays an important role in a number of physics analysis scenarios at the LHC, particularly with regard to searches for new particles beyond the Standard Model, which is the currently-accepted model of particle interactions. Notable examples are the search for Dark Matter candidates, i.e. for new particles that could explain $\sim$85\% of the mass of the universe currently not accounted for, as well as searches for new particles predicted by the theory of supersymmetry and by theories that postulate the existence of extra dimensions. Moreover, missing transverse energy is an essential ingredient in the study of a number of Standard Model processes, such as the decay of the recently-discovered Higgs-like boson to pairs of $\tau$ leptons, as well as processes that involve $W$ bosons and top quarks in the final state. 

In our previous studies, we proposed the idea of filtering individual events particle by particle at the reconstruction level in order to improve on the rejection of contamination from low-energy strong interactions in high-luminosity hadron collider environments. The algorithm that we describe in this article is a simplified deterministic variant of the Markov Chain Monte Carlo technique that we used in \cite{gibbshep2,gibbshep}. 

It is our opinion that the simplicity and parallelisation potential of this technique make it a promising candidate for inclusion in particle-by-particle event filtering procedures at the reconstruction level at future high-luminosity hadron collider experiments.

We see this algorithm as complementary to the particle weighting methods that have been recently proposed at the LHC. Since our technique is based on a different approach, we expect its combination with state-of-the-art algorithms to result in improved performance at higher pileup rates. As more particle weighting methods are proposed, we also envisage the possibility of combining the different weights, e.g. in the context of a multivariate framework, with a view to exploiting all the information available in the data at the level of individual particles.

\section{Background\label{background}}

Historically, methods of subtracting contamination from soft QCD interactions have been developed with a view to correcting observables associated with hard jets, i.e. with collections of final-state particles produced by the showering and hadronisation of scattered high-energy partons.

The state of the art includes a number of techniques that are often based on different principles, and that are typically used in combination at the ATLAS and CMS experiments at the LHC. 

In this context, an important role is played by correction procedures that relate to the concept of jet area \cite{jet-area}, which provides a measure of the susceptibility of jets to contamination from low-energy particles. A core ingredient of jet area-based algorithms is the estimation of an event-level soft QCD transverse momentum\footnote{
The transverse momentum, $p_T$, of a particle is defined as the absolute value of the component of the particle momentum vector on a plane perpendicular to the direction of the colliding beams.
}
 density based on the pileup jets in the event. With such methods, which are core ingredients of the official reconstruction and analysis pipelines at the LHC, the correction applied to the total transverse momentum of a hard jet of interest is proportional both to the jet area and to an estimate of the soft QCD transverse momentum density in the event. However, techniques based on jet area are not designed to describe differences in soft QCD energy flow between different kinematic regions inside collision events, and the upcoming high-luminosity regimes are likely to call for the development of dedicated methods.

With the introduction of jet substructure techniques, soft QCD contamination started to be studied in terms of individual components inside hard jets, thereby exploiting the hierarchical structure of jets. This has resulted in an important suite of new tools for reconstruction and analysis at the LHC, particularly with regard to jet grooming \cite{jet-filtering,jet-trimming,jet-pruning-1,jet-pruning-2,jet-soft-drop} and jet cleansing \cite{jet-cleansing}. 

In general terms, the algorithmic evolution outlined above has gradually moved toward more ``local'' estimates of soft QCD contamination that take into account the variability across collisions as well as inhomogeneity inside individual events. This has ultimately led to the development of methods working at the level of individual particles as the most fine-grained level of information available in the data. Notable examples are PUPPI \cite{REVIEW_PILEUP_2014,PUPPI}, SoftKiller \cite{SoftKiller} and the particle-level technique presented in \cite{berta}. 

For instance, PUPPI exploits the existence of collinear singularities in the physics that underlies the showering process. This makes it possible to assign individual particles weights that reflect the likelihood of them originating from the hard parton scattering as opposed to soft QCD interactions. Specifically, the weights rely on a measure of proximity between particles in a space defined in terms of particle transverse momentum, $p_T$, pseudorapidity\footnote{
Particle pseudorapidity, $\eta$ is a kinematic quantity expressed in terms of the particle polar angle in the laboratory frame by $\eta=-\mbox{log}\left[\mbox{tan}(\theta/2)\right]$.
}, $\eta$, and azimuthal angle, $\varphi$. 

\section{The approach{\label{method}}}

The probability density functions (PDFs) that describe the kinematics of particles originating from soft QCD interactions as opposed to a hard parton scattering reflect the properties of the underlying physics processes, and describe the expected shapes of the corresponding particle-level distributions. However, even when the processes involved are exactly the same, individual collision events contain independent, and therefore different, realisations of the underlying quantum processes, and the shapes of the corresponding particle-level distributions are for this reason generally different in different events. In other words, the shape of the kinematic distribution of particles originating from soft QCD interactions is generally event-specific, i.e. each event can in principle be associated with its own particle-level soft QCD kinematic ``fingerprint''. 

A key aspect of our approach is the idea of using the particle weights to estimate the shape of the soft QCD kinematic distribution in terms of particle $\eta$ and $p_T$ inside individual events, thereby taking into account the inherent variability across collisions due to the presence of statistical fluctuations in the data. Given an estimate of the neutral soft QCD particle fraction in each event, this is equivalent to estimating the corresponding number of neutral soft QCD particles in different $(\eta, p_T)$ bins.

Although the actual numbers of particles originating from background soft QCD interactions as opposed to the signal hard scattering are not known, given a signal model, it is possible to estimate the expected number of signal particles, $\nu_s(\eta, p_T)$, in each $(\eta, p_T)$ region. On the other hand, the expected number of background particles normally cannot be estimated due to the non-perturbative nature of the underlying physics processes.

If $n_s^*(\eta, p_T)$ and $n_b^*(\eta, p_T)$ denote the unknown true numbers of signal and background soft QCD particles in each $(\eta, p_T)$ bin, then $n(\eta, p_T) = n_s^*(\eta, p_T) + n_b^*(\eta, p_T)$, where $n(\eta, p_T)$ is the corresponding number of particles in the data. In general, whenever an event exhibits an excess\footnote{
A similar line of reasoning applies to a depletion in the number of particles.
} of particles in a given $(\eta, p_T)$ region as compared to the average number, it is not known to what extent the excess originates from a soft QCD interaction as opposed to the hard scattering. However, when one considers LHC events with a number of vertices in line with what is expected in the upcoming higher-luminosity regimes, the final-state particle multiplicities associated with soft QCD background interactions and with the signal hard scattering are such that the expected number of signal particles is typically much lower than the number of background particles, i.e.

\begin{equation}
\left<\nu_s(\eta, p_T)\right>\ll \left<n_b^*(\eta, p_T)\right>,
\end{equation}

where the average is taken over the $(\eta, p_T)$ space. This also implies that the statistical fluctuations on the number of particles in a given $(\eta, p_T)$ region are typically dominated by the fluctuations on the number of soft QCD particles, i.e. $\left<\sigma_{n_s}(\eta, p_T)\right> \ll \left<\sigma_{n_b}(\eta, p_T)\right>$. Under such conditions, it is reasonable to express the estimated number of soft QCD particles in terms of

\begin{equation}
\hat{n}_b(\eta, p_T) \simeq n(\eta, p_T)-\nu_s(\eta, p_T)
\label{eq:sigmanb}
\end{equation}

In the following, we estimate the shape of the particle-level $(\eta, p_T)$ distribution of neutral soft QCD particles inside individual events using an event-level estimate of the neutral soft QCD particle fraction, as well as PDF templates obtained from high-statistics control samples. The procedure is outlined below:

\begin{enumerate}

\item Control samples are first used to estimate the shapes of the expected ($\eta$, $p_T$) distributions of neutral final-state particles originating from soft QCD interactions and from the signal hard parton scattering. Such distributions reflect the properties of the underlying physics processes, and their shapes correspond to what is expected from an average over multiple events.

\item The overall fraction of neutral soft QCD particles in each event is estimated based on the corresponding charged particle fraction. 

\item The above information is used to define weights that reflect the probability for individual particles to originate from soft QCD interactions as opposed to the hard scattering.

\item The weights are employed to reshape the particle-level ($\eta$, $p_T$) distribution in the data, with a view to estimating the number of neutral soft QCD particles in different kinematic regions event by event.

\end{enumerate}

\section{The algorithm\label{algo}}

For the purpose of this study, the particle-level $(\eta, p_T)$ space is each event has been subdivided into bins of widths $\Delta\eta=0.5$ and $\Delta p_T=0.05$~GeV/c. We focus on particles with $0<p_T<1$~GeV/c, which are the majority of those produced by soft QCD interactions. The algorithm consists of the following steps, along the lines discussed in the previous section:

\begin{enumerate}[label=\arabic*]

\item Obtain the shapes of the particle-level $(\eta, p_T)$ PDFs from the high-statistics control samples. In the following, $f_0(\eta, p_T)$ and $f_1(\eta, p_T)$ will denote the PDFs of neutral particles originating from soft QCD interactions and from the signal hard scattering, respectively.

\item In each event, estimate the overall fraction of neutral soft QCD particles, $\alpha_0^{(n)}$, in terms of the corresponding charged particle fraction, $\alpha_0^{(c)}$:

\begin{equation}
\hat{\alpha}_0^{(n)}=\mbox{min}(k \alpha_0^{(c)}, \alpha_0^{(c)})
\label{eq:alpha_0}
\end{equation}

The role of the correction factor $k$, which is estimated from Monte Carlo as described in section \ref{results}, is to correct on average for the difference between neutral and charged particle kinematics. This includes a correction for the number of charged particles with $p_T$ below 500~MeV/c that do not reach the tracking detectors. Taking the minimum in (\ref{eq:alpha_0}) ensures that $\hat{\alpha}_0^{(n)}$ is always lower than 1.

\item Combine the above information into particle weights:

\begin{equation}
w_0(\eta, p_T) = \displaystyle \frac{\hat{\alpha}_0^{(n)} f_0(\eta, p_T)}{\hat{\alpha}_0^{(n)} f_0(\eta, p_T) + \hat{\alpha}_1^{(n)} f_1(\eta, p_T)},
\label{eq:w_0}
\end{equation}

with $\alpha_0^{(n)} + \alpha_1^{(n)} = 1$. The quantity $w_0(\eta, p_T)$ provides an estimate of the probability for individual particles in each $(\eta, p_T)$ bin to originate from soft QCD interactions as opposed to the hard parton scattering of  interest.

\item Use $w_0(\eta, p_T)$ to reshape the $(\eta, p_T)$ distribution of neutral particles in the data in order to estimate the distribution of neutral soft QCD particles in each event. The expected number of neutral soft QCD particles, $\hat{n}_b(\eta, p_T)$, is estimated in terms of

\begin{equation}
\hat{n}_b(\eta, p_T) = w_0(\eta, p_T) n(\eta, p_T),
\label{eq:sculpt}
\end{equation}

where $n(\eta, p_T)$ is the corresponding number of neutral particles in the data. Given the expected number $\hat{n}_b(\eta,p_T)$, the unknown number of neutral soft QCD particles in each $(\eta, p_T)$ bin can be treated as a random variable following a binomial distribution with mean given by (\ref{eq:sculpt}) and standard deviation

\begin{equation}
\sigma_{\hat{n}_b} = \sqrt{n w_0 (1-w_0)}.
\label{eq:dnsculpt}
\end{equation}

It should be noted that (\ref{eq:sculpt}) is equivalent to (\ref{eq:sigmanb}) if one uses $w_0$ to calculate $\nu_s$, i.e. if $\nu_s/n = 1-w_0$.

\item Estimate the number of neutral soft QCD particles in each $(\eta, p_T)$ bin in terms of:

\begin{equation}
\hat{n}_b = w_0 n \pm \sqrt{n w_0 (1-w_0)}
\end{equation}

\end{enumerate}

It is worth noticing that the algorithm is inherently parallel, since different bins can be processed independently. It is our opinion that the simplicity and parallelisation potential of this technique make it a promising candidate for inclusion in future particle-level event filtering procedures upsteam of jet reconstruction at high-luminosity hadron collider experiments.

In this article, we have used the weights defined in (\ref{eq:w_0}) to illustrate this approach. However, it should be emphasised that the idea of employing the weights to reshape the particle-level $(\eta, p_T)$ distribution inside individual events does not require this choice of weights, and can in principle be used in conjunction with any particle weighting procedure, as discussed in section \ref{weights}.

\begin{figure*}
\centering
\subfloat[]{
\includegraphics[scale=0.37]{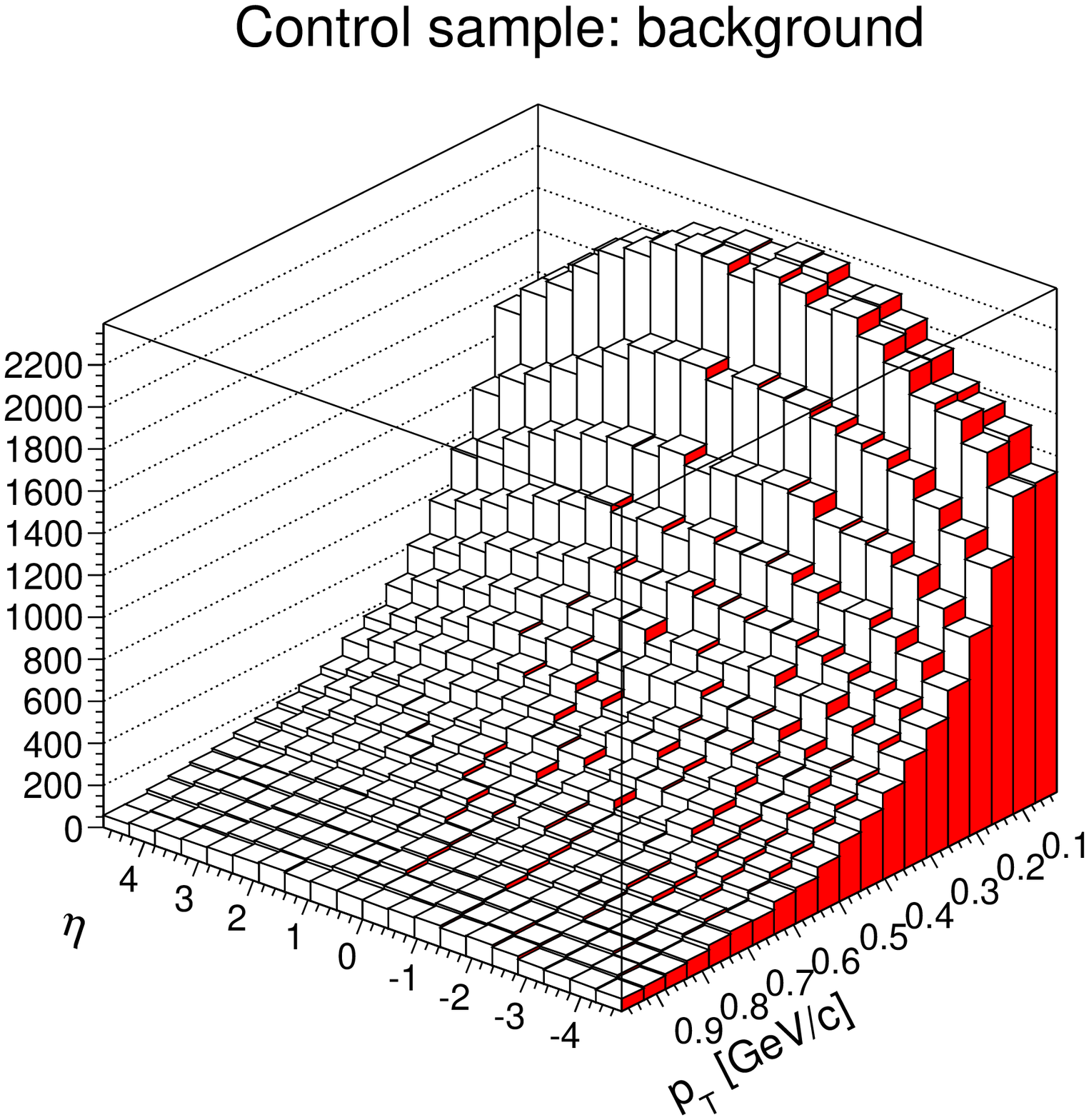}
}
\subfloat[]{
\includegraphics[scale=0.37]{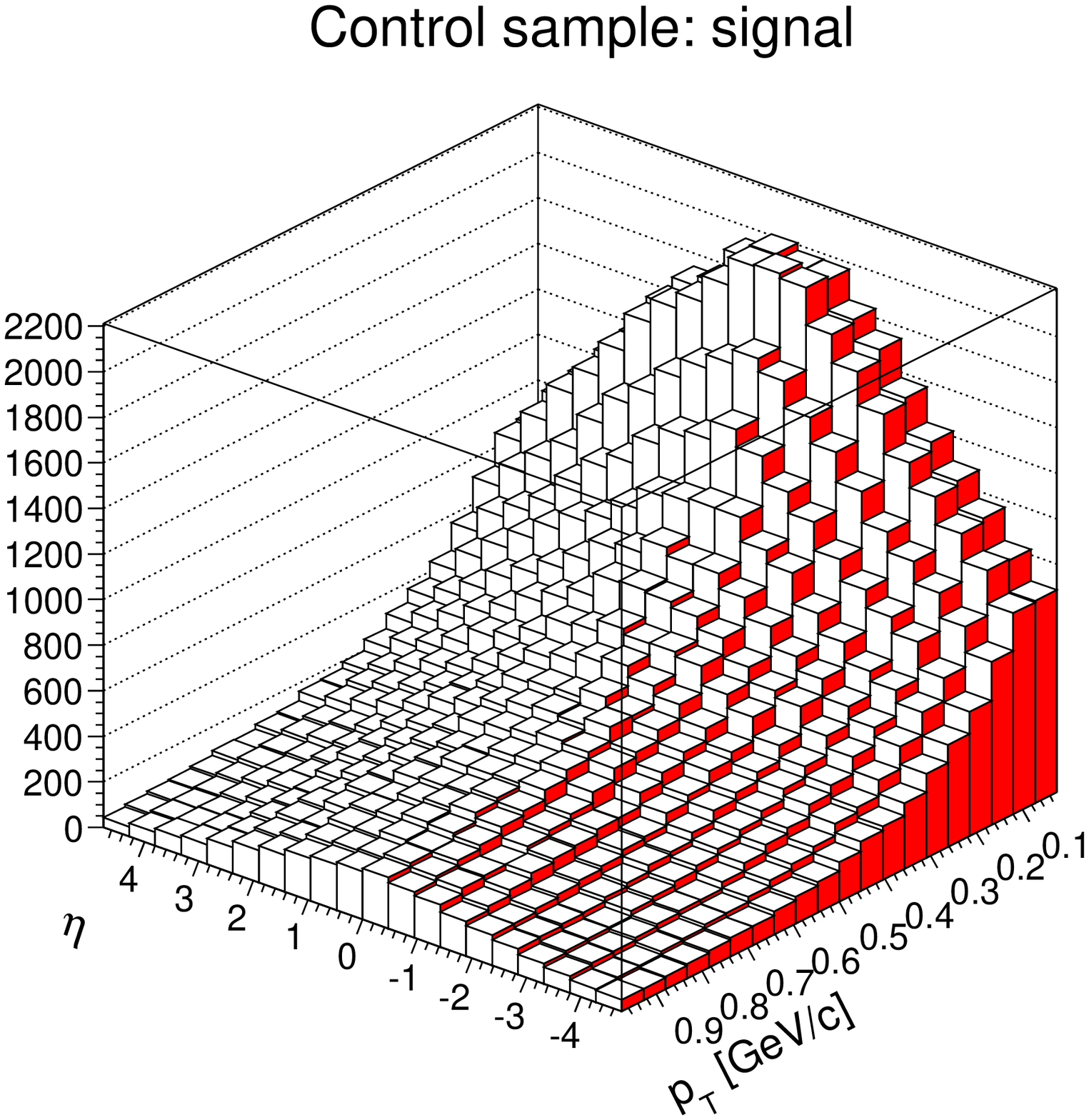}
}
\caption[]{
Particle-level $(\eta, p_T)$ distributions from the high-statistics control samples in the region $-5<\eta<5$, $0<p_T<1~\mbox{GeV/c}$, as described in the text. (a) Neutral soft QCD particles. (b) Neutral particles from the hard parton scattering. The plots have been rotated around the $z$ axis in order to make the kinematic signatures more clearly visible.
}
\label{fig:cs}
\end{figure*}

\section{Results\label{results}}

We discuss the results of a feasibility study of this approach on Monte Carlo data at the generator level. We used Pythia 8.176 \cite{pythia1,pythia2} to generate 1,000 events, each consisting of a $gg\rightarrow{t\bar{t}}$ hard parton scattering at $\sqrt{s}=14$~TeV superimposed with 50 soft QCD interactions to simulate the presence of pileup.

\subsection{Control sample PDF templates}

We generated control sample data sets containing $\sim300,000$ particles originating from the signal $gg\rightarrow{t\bar{t}}$ hard scattering and $\sim300,000$ particles associated with background soft QCD interactions. These data sets were used to obtain the shapes of the $(\eta, p_T)$ PDFs of signal and background neutral particles. The latter reflect the particle-level kinematic signatures of the underlying physics processes, from which the corresponding distributions in the data generally deviate due to the presence of statistical fluctuations.

Figure \ref{fig:cs} displays the corresponding $(\eta, p_T)$ distributions of neutral soft QCD particles (a) and of neutral particles from the hard scattering (b), each normalised to unit volume.

\subsection{Event-by-event neutral particle fractions}
\label{alpha}

One of the pieces of information required by the choice of weights that we have made for the purpose of this study is an event-by-event estimate of the overall fraction of neutral particles originating from soft QCD interactions.

We estimate the neutral pileup particle fraction in each event in terms of the corresponding charged fraction. We apply a correction factor, $k$, that represents an average over multiple Monte Carlo events according to (\ref{eq:alpha_0}), i.e $\hat{\alpha}_0^{(n)} = \mbox{min}(k \hat{\alpha}_0^{(c)}, \alpha_0^{(c)})$, where $\alpha_0^{(n)}$ and $\alpha_0^{(c)}$ are the overall neutral and charged pileup particle fractions in the event, respectively. Specifically, the correction factor is given by $k=<\alpha_0^{MC,n}/\alpha_0^{MC,c}>$, where $\alpha_0^{MC,n}$ ($\alpha_0^{MC,c}$) is the fraction of neutral (charged) pileup particles estimated from Monte Carlo, and the average is taken over the 1,000 events generated in this study.

\begin{figure*}
\centering
\begin{minipage}{17pc}
\includegraphics[scale=0.37]{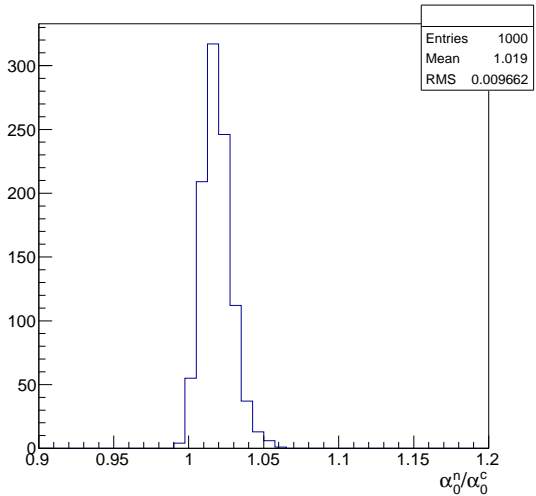}
\end{minipage}\hspace{3pc}%
\begin{minipage}{14pc}
\caption{\label{fig:fracs}Distribution of the ratio between the fraction of neutral soft QCD particles and the corresponding fraction of charged particles, $\alpha_0^{MC,n}/\alpha_0^{MC,c}$ from Monte Carlo, over the events generated in this study. Additional information is given in the text.}
\end{minipage}
\end{figure*}

Figure \ref{fig:fracs} displays the ratio between the fraction of neutral pileup particles and the corresponding quantity for charged particles in the events generated. The results shown in the following have been obtained using $k=1.02$, which corresponds to the mean of the distribution in figure \ref{fig:fracs}. Multiple runs of the algorithm were performed whereby $k$ was varied within 5\% of its nominal value, and produced consistent results.

\subsection{Particle weights}
\label{weights}

For the purpose of this study, the particle weights have been defined as a function of the fraction of neutral soft QCD particles in each event as well as of the control sample PDF templates, according to (\ref{eq:w_0}). To our knowledge, this is the first particle weighting method that directly exploits the particle-level kinematic signatures of the underlying physics processes in the $(\eta, p_T)$ space. We see this choice of weights as complementary to those adopted in the recently-proposed algorithms that use measures of proximity between particles defined in terms of particle $p_T$, $\eta$ and $\varphi$.

According to the above choice of weights, all particles in the same $(\eta, p_T)$ bin are assigned the same weight. However, as previously noticed, the idea of employing the weights to reshape the distribution in the data is more general and can in principle be used in conjunction with any particle weighting method. In fact, if $S(\eta, p_T)$ denotes the set of particles in each $(\eta, p_T)$ bin in the data and $n(\eta, p_T)$ the corresponding particle multiplicity, rescaling $n(\eta, p_T)$ by $\sum_{i\in S(\eta, p_T)} w_i / n(\eta, p_T)$ reduces to (\ref{eq:sculpt}) when $w_i\equiv w_0(\eta, p_T)$.

In other words, rescaling the $(\eta, p_T)$ distribution in the data in order to estimate the number of soft QCD particles across the particle kinematic space is equivalent to setting the bin contents to a function of the data that is given by $\sum_{i\in S(\eta, p_T)} w_i$. If it were known which particles in the event originate from the hard scattering and which from soft QCD interactions, the weights would be either 0 or 1. With reference to those $(\eta, p_T)$ bins that contain no signal particles, i.e. where all particles in the bin originate from soft QCD interactions, $w_i\equiv w_0(\eta, p_T) = 1$ and $\sum_{i\in S(\eta, p_T)} w_i = n_b^*(\eta, p_T)$. With regard to those bins, the problem of estimating the number of soft QCD particles becomes trivial, and $\hat{n}_b(\eta, p_T) = n(\eta, p_T)$. Correspondingly, the uncertainty associated with the estimated number of soft QCD particles, according to (\ref{eq:dnsculpt}), becomes zero.

In reality, it is only possible to estimate a probability for individual particles to originate from either process, also in the light of the role played by colour connection, and $w_i\in [0,1]$. The results presented in the following show that our method produces reasonable estimates of the number of neutral soft QCD particles in different $(\eta, p_T)$ bins regardless of whether or not particles originating from the signal hard scattering are present. In general, the accuracy of $\hat{n}_b(\eta, p_T)$ is expected to increase with increasing accuracy of the weights. 

It is worth noticing that the use of $w_0(\eta, p_T)$ is associated with a relatively coarse-grained decomposition of the $(\eta, p_T)$ space, particularly along the $\eta$ axis. However, as mentioned above, we are not proposing to use these weights in isolation, but rather in combination with other particle-level metrics, such as those presented in \cite{PUPPI}.

It should also be emphasised that $w_0(\eta, p_T)$ encodes properties of the underlying physics processes that are not employed by other methods, and we expect the combined use of different weights to be beneficial. For instance, some of the results presented in \cite{PUPPI} seem to suggest over-subtraction of soft QCD particles, whereby particles originating from the hard scattering are erroneously misidentified as pileup-related. We foresee the possibility of implementing optimised particle weighting algorithms that make use of all the particle-level information available in the data, thereby further improving on the performance of pileup subtraction in high-luminosity environments.

\subsection{Soft QCD kinematic fingerprints}
\label{fingerprints}

A key aspect of our approach is the use of bin-by-bin estimates of the neutral soft QCD particle fractions in order to estimate how the corresponding particles are distributed across the $(\eta, p_T)$ space in each event. This method therefore takes into account the variability of the shape of the soft QCD particle distribution across collisions, i.e. it enables the estimation of event-specific soft QCD kinematic ``fingerprints'' at the particle level.

The performance of the algorithm is illustrated in the following with regard to one of the Monte Carlo events generated in this study, chosen as a reference. Consistent results were obtained on all events analysed.

\begin{figure*}
\centering
\subfloat[]{
\includegraphics[scale=0.37]{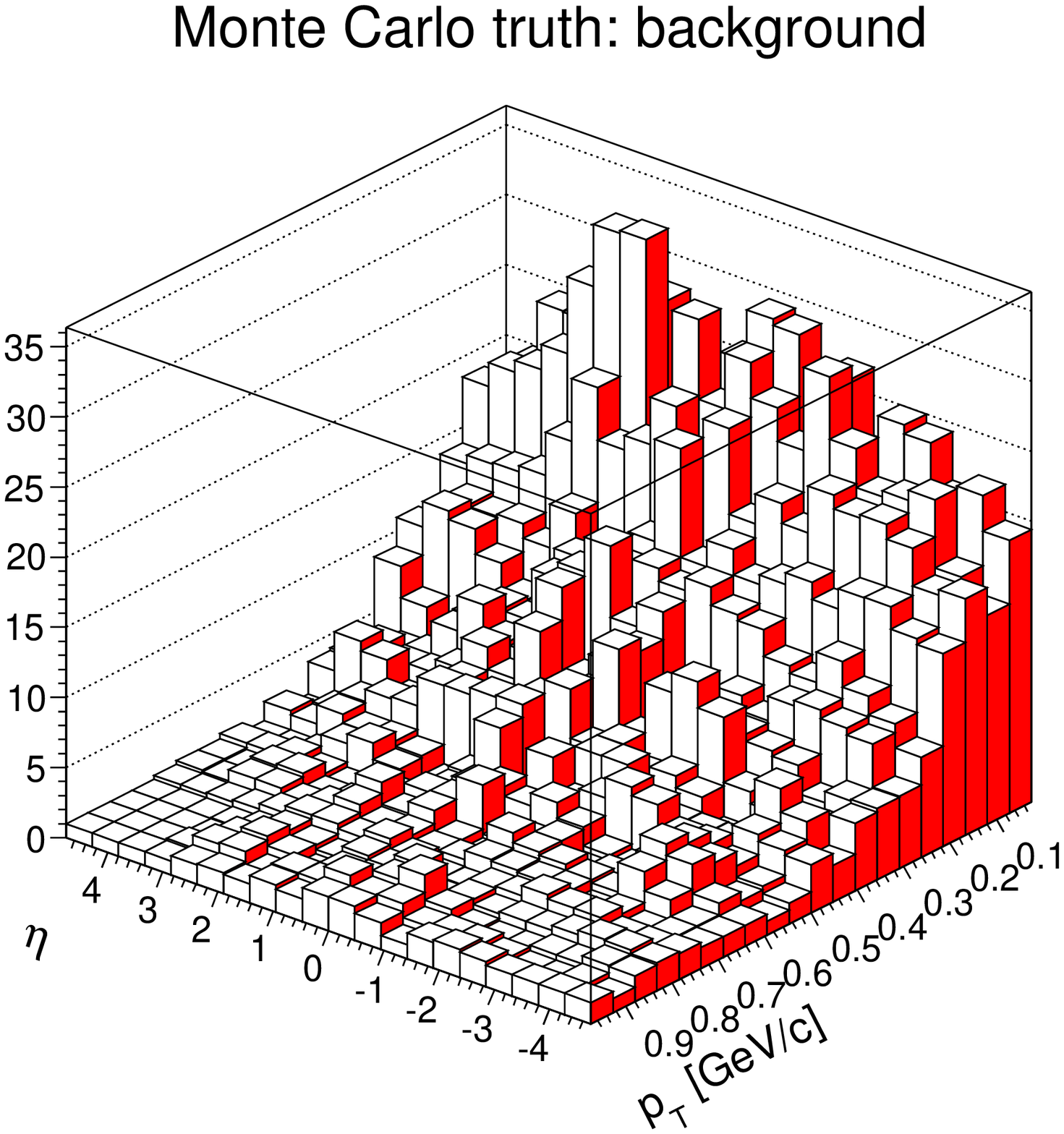}
}
\subfloat[]{
\includegraphics[scale=0.37]{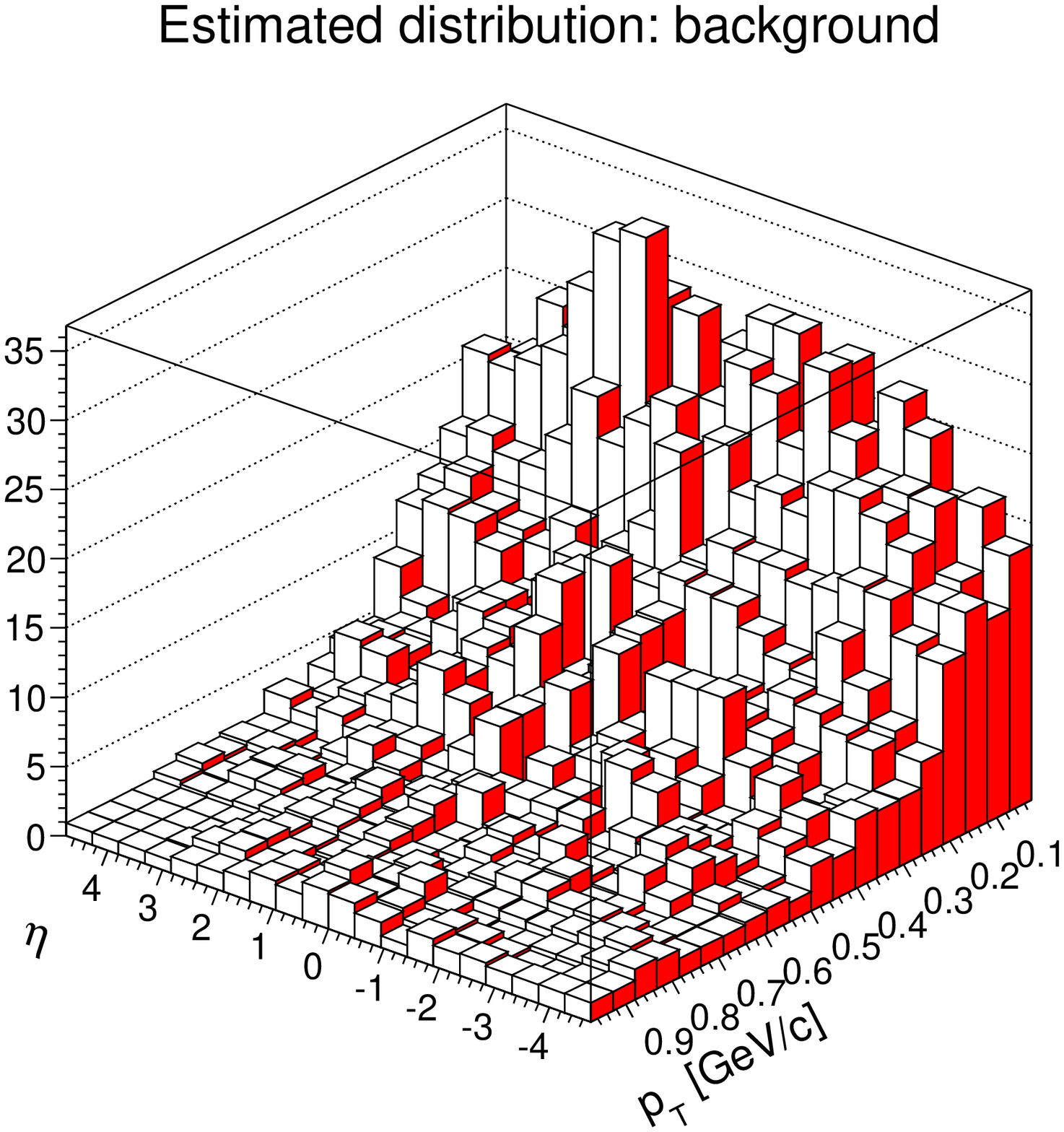}
}
\caption[]{
(a) True particle-level $(\eta, p_T)$ distribution of neutral soft QCD particles from one of the Monte Carlo events generated in this study. The plot highlights the deviation of the shape of the distribution from that of the corresponding control sample template due to the presence of statistical fluctuations in the data. (b) The corresponding particle-level $(\eta, p_T)$ distribution of neutral soft QCD particles estimated using this algorithm as detailed in the text.
}
\label{fig:truth}
\end{figure*}

Figure \ref{fig:truth} (a) displays the true particle-level ($\eta$, $p_T$) distribution of neutral soft QCD particles in the reference event. As shown by a comparison with the corresponding high-statistics control sample distribution in figure \ref{fig:cs} (a), the particle-level $(\eta, p_T)$ soft QCD distribution inside individual events generally exhibits local features that are washed out when multiple events are lumped together.

\subsection{Soft QCD particle counting{\label{count}}}

We illustrate here the proposed use of particle weights to reshape the particle-level $(\eta, p_T)$ distribution in the data, using the weights defined in section \ref{weights} as an example. The objective is to estimate $\hat{n}_b(\eta, p_T)$, which describes how soft QCD neutral particles are distributed across the $(\eta, p_T)$ space inside individual events. A tentative estimate could in principle be obtained using the control sample $(\eta, p_T)$ PDF templates, $f_0$ and $f_1$, in terms of 

\begin{equation}
N^{(n)}\displaystyle\frac{\alpha_0^{(n)}f_0(\eta, p_T)}{\alpha_0^{(n)}f_0(\eta, p_T)+\alpha_0^{(n)}f_1(\eta, p_T)}, 
\label{eq:guess}
\end{equation}

where $N^{(n)}$ is the total number of neutral particles in the event. However, although this relies on an event-by-event estimate of the neutral soft QCD particle fraction, it cannot account for the inhomogeneity of the distribution of soft QCD particles that is typically observed inside individual events.

If the unknown true probability densities for a particle with transverse momentum $p_T$ and pseudorapidity $\eta$ to originate from the signal hard scattering and from background soft QCD interactions are denoted by $f^*_1(\eta, p_T)$ and $f^*_0(\eta, p_T)$, the true fractions of background and signal particles inside each $(\eta, p_T)$ bin are $f^*_0(\eta, p_T)\Delta\eta\Delta p_T$ and $f^*_1(\eta, p_T)\Delta\eta\Delta p_T$, respectively, where $\Delta\eta$ and $\Delta p_T$ are the widths of the bins along the $\eta$ and $p_T$ axes.

In reality, $f_0^*$ and $f_1^*$ are not known, and we use the corresponding control sample PDF templates in (\ref{eq:w_0}) to estimate the local fractions of soft QCD particles. 

If the unknown true number of soft QCD particles, $n_b^*(\eta, p_T)$, is higher than the corresponding average number due to fluctuations in the data, a fraction $w_0(\eta, p_T)$ of the excess will add to the estimated number of neutral soft QCD particles in that $(\eta, p_T)$ bin. At the same time, a fraction $1-w_0(\eta, p_T)$ will be associated with the hard parton scattering. A similar line of reasoning can be applied to a situation whereby fluctuations lead to a depletion in terms of the number of soft QCD particles with respect to the average. In other words, $\hat{n}_b(\eta, p_T)$ is generally expected to reflect the true distribution $n^*_b(\eta, p_T)$ more accurately than (\ref{eq:guess}).

While $\hat{n}_b(\eta, p_T)$ represents the expected number of neutral soft QCD particles, the unknown actual number can be treated as a random variable following a binomial distribution with mean $n(\eta, p_T) w_0(\eta, p_T)$ and standard deviation given by (\ref{eq:dnsculpt}). The estimated $(\eta, p_T)$ distribution of neutral soft QCD particles in the event chosen to illustrate our method is shown in figure \ref{fig:truth} (b). As can be seen, the local features of the distribution due to the presence of fluctuations in the data are reasonably described, e.g. the excess at $\eta\simeq 2.5$ and $p_T\simeq 0.2$~GeV/c.

\begin{figure*}
\centering
\subfloat[]{
\includegraphics[scale=0.37]{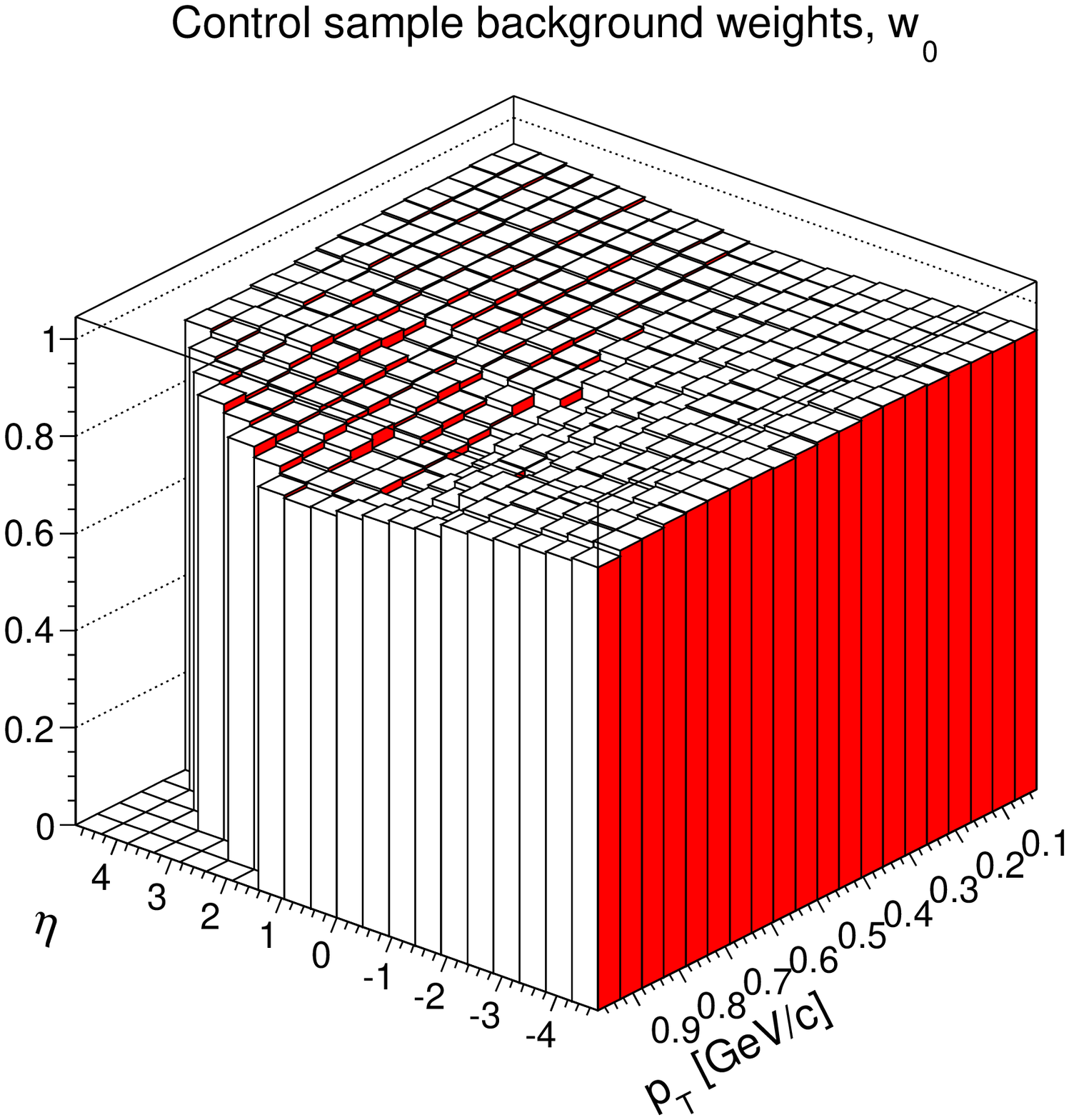}
}
\subfloat[]{
\includegraphics[scale=0.37]{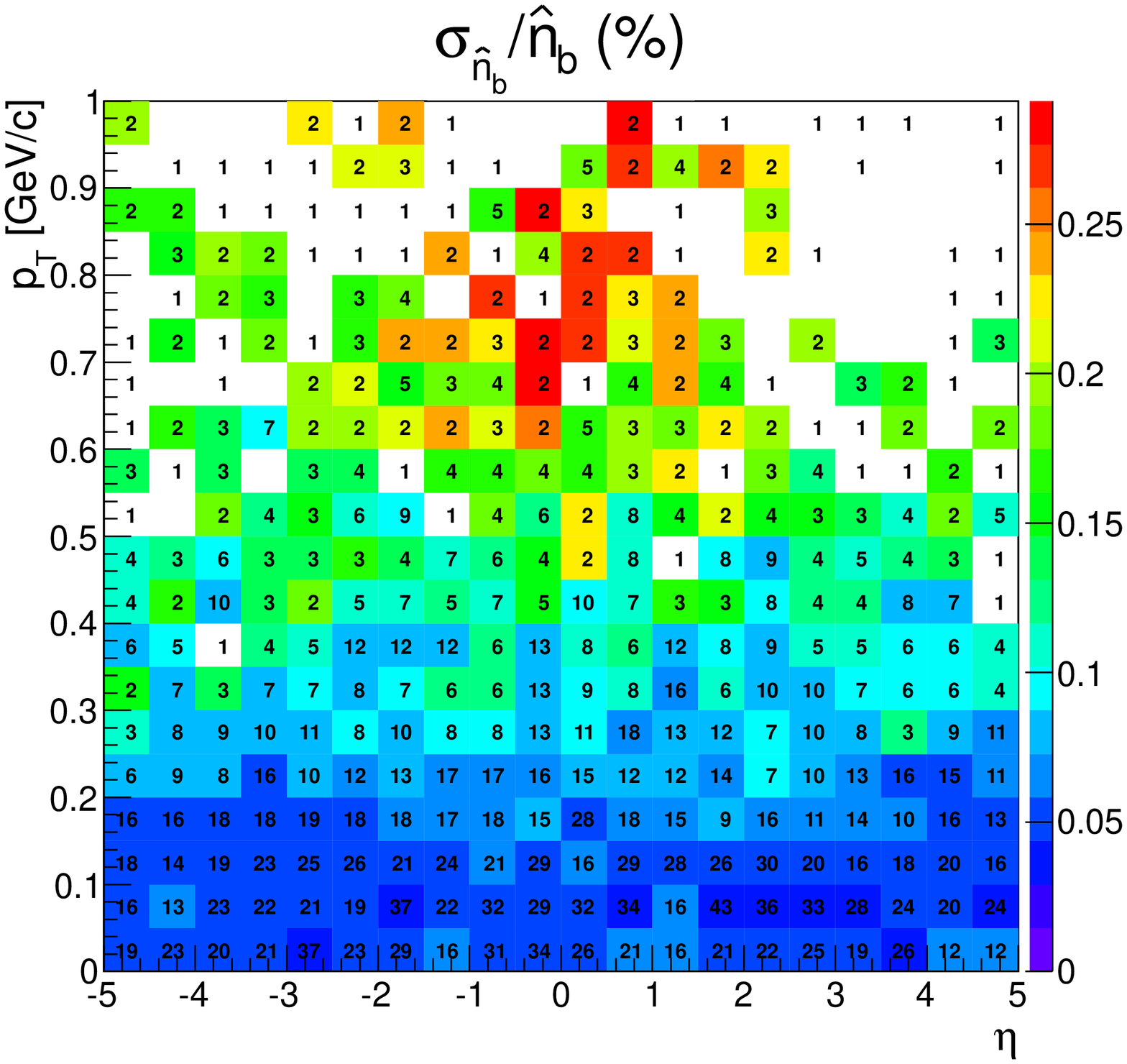}
}
\caption[]{
(a) Particle weights as defined in the text. (b) Relative uncertainty on the estimated number of neutral soft QCD particles, $\sigma_{\hat{n}_b}/\hat{n}_b$, represented as a heatmap superimposed to $\hat{n}_b(\eta, p_T)$. Additional information is provided in the text.
}
\label{fig:heatmap}
\end{figure*}

The same results are shown in figure \ref{fig:heatmap}, where the estimated number of soft QCD particles across the $(\eta, p_T)$ space is superimposed with a heatmap corresponding to the relative uncertainty on $\hat{n}_b(\eta, p_T)$. As expected, the uncertainty is higher in lower-statistics bins, i.e. $w_0$ is a more reliable estimate of the local fraction of soft QCD particles in more highly-populated bins.

Based on these results, we expect the use of this technique in conjunction with the recently-proposed particle weighting methods to be particularly beneficial in the region $p_T < 500$~MeV/c, which contains most of the particles originating from soft QCD interactions.

\begin{figure*}
\centering
\subfloat[]{
\includegraphics[scale=0.17]{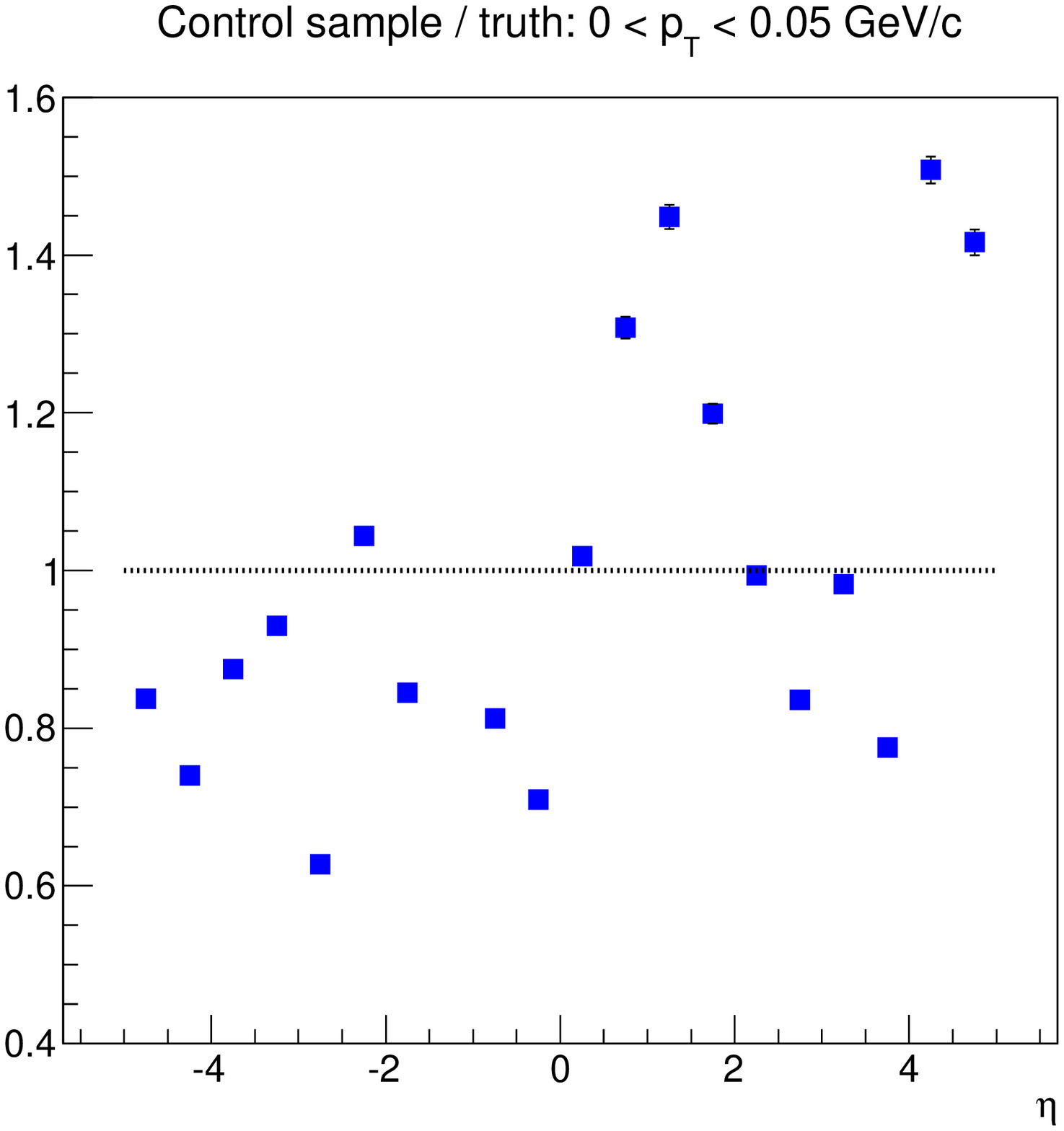}
}
\subfloat[]{
\includegraphics[scale=0.17]{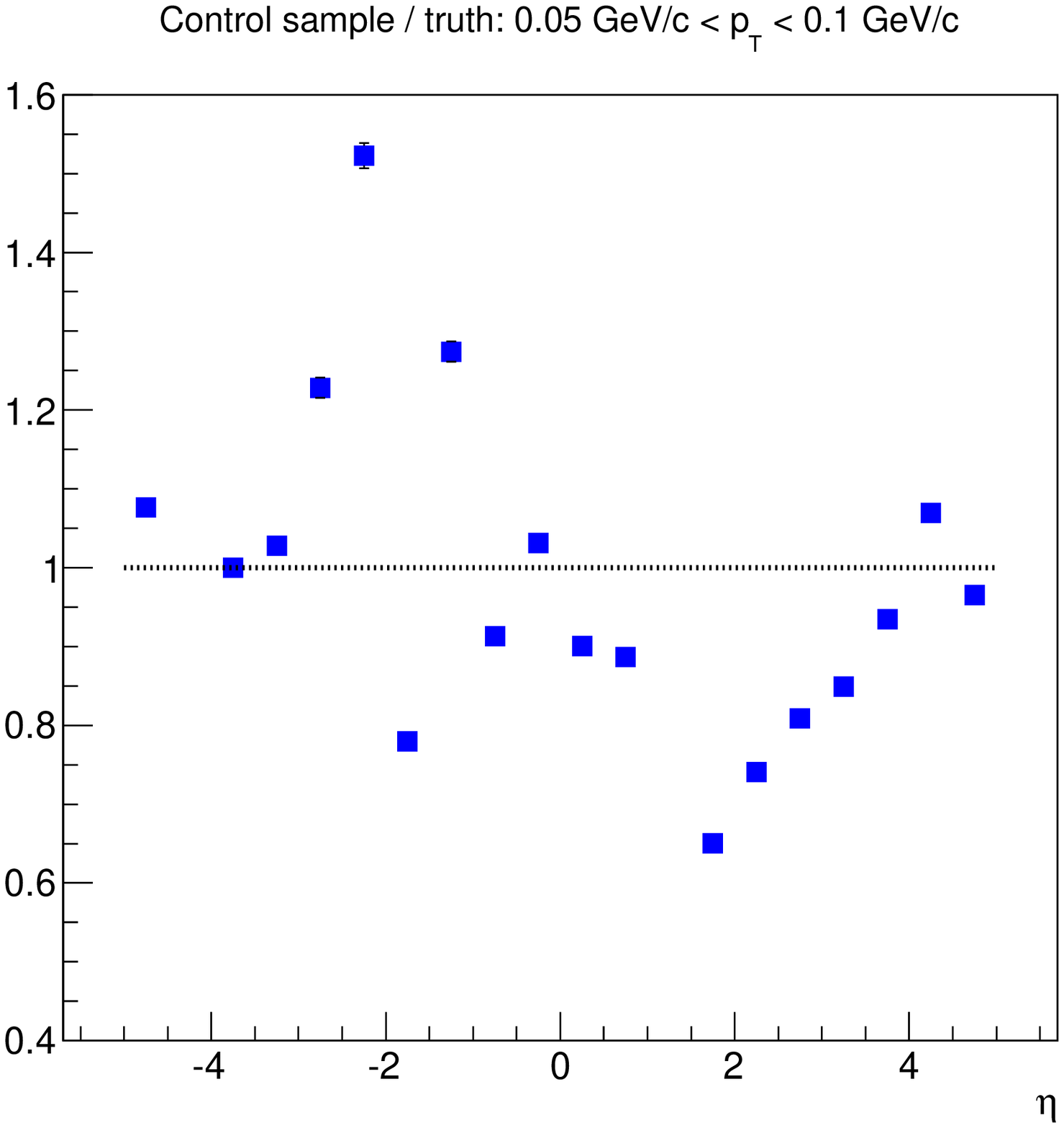}
}
\subfloat[]{
\includegraphics[scale=0.17]{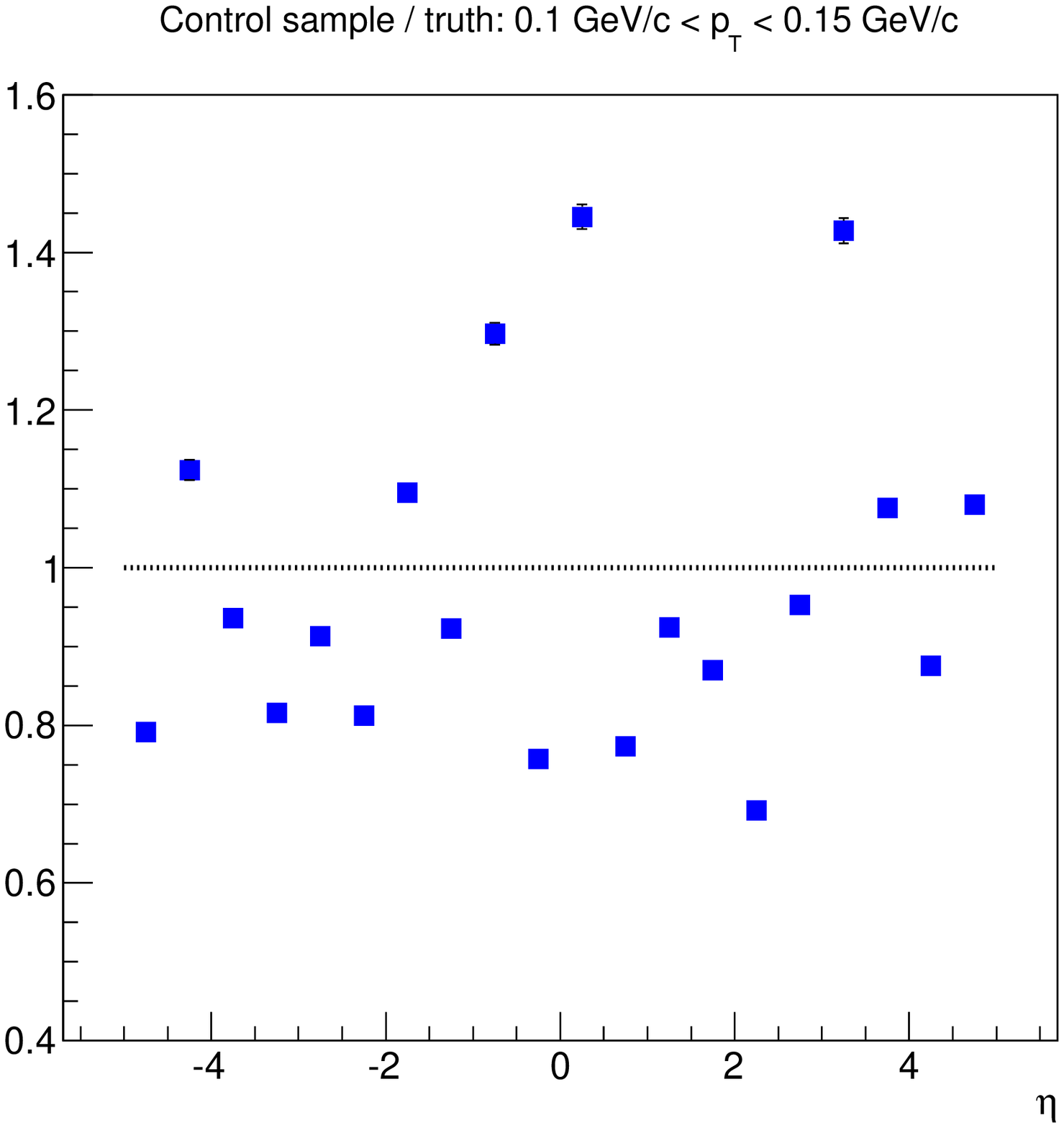}
}
\subfloat[]{
\includegraphics[scale=0.17]{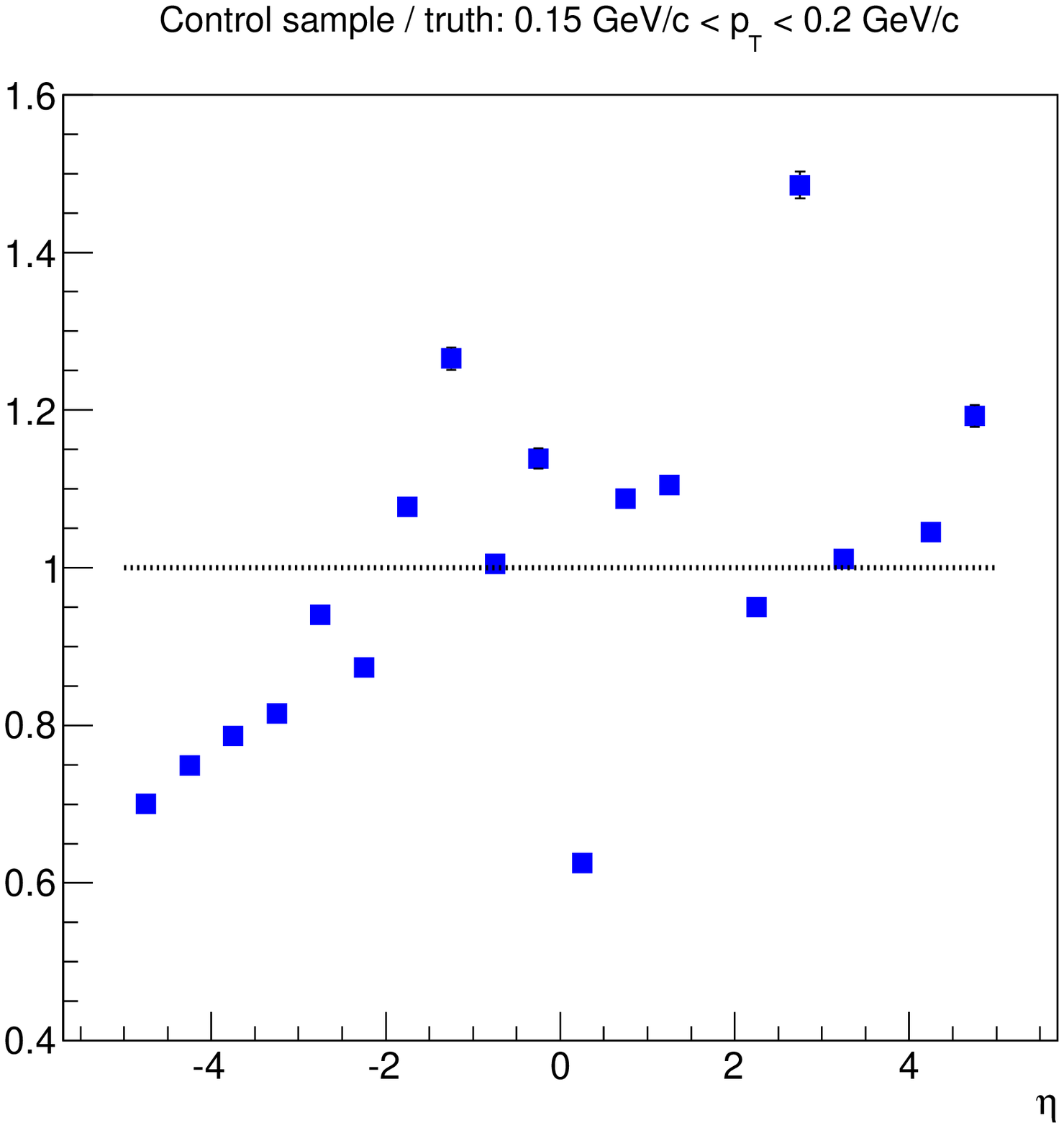}
}\\
\subfloat[]{
\includegraphics[scale=0.17]{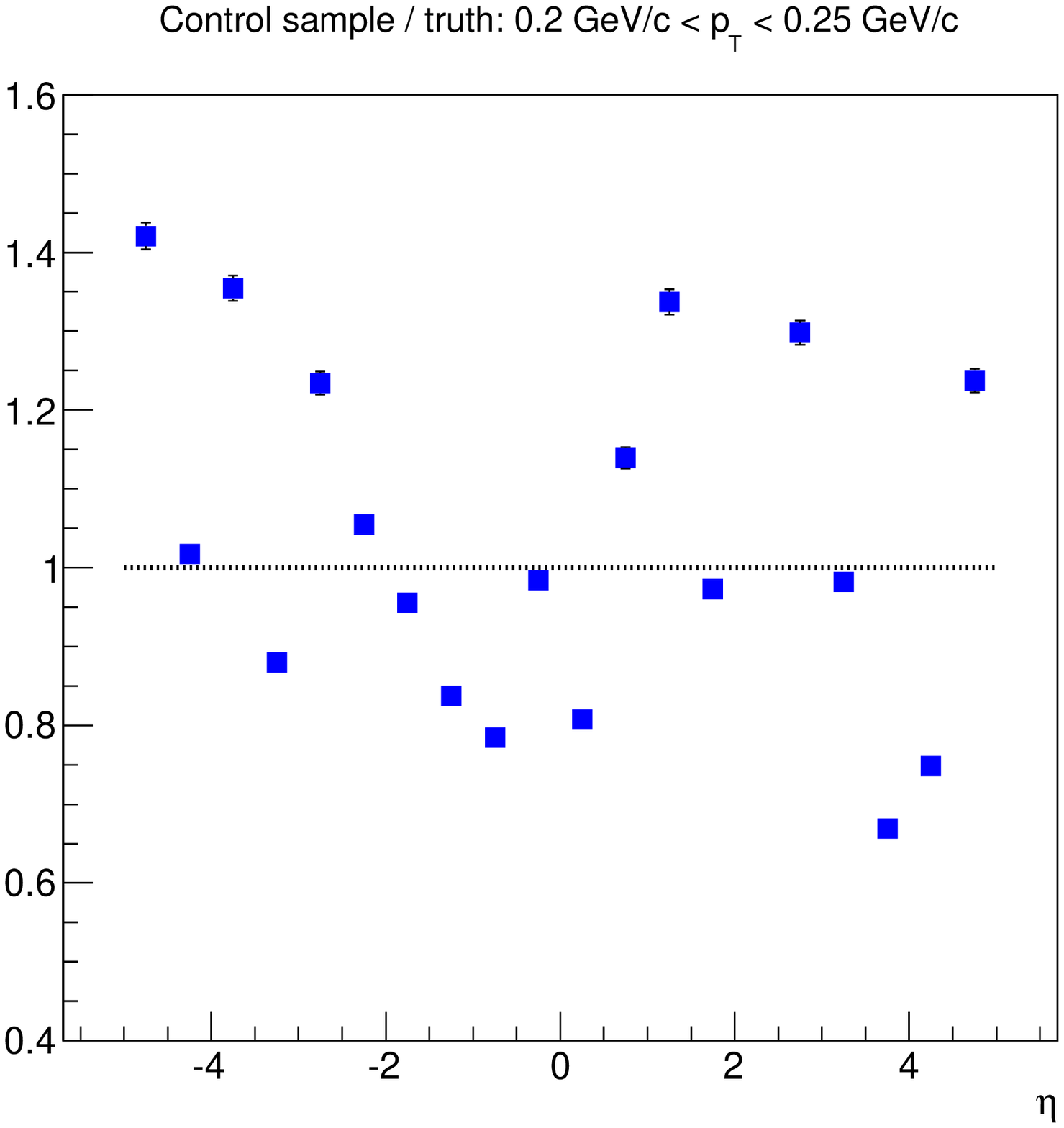}
}
\subfloat[]{
\includegraphics[scale=0.17]{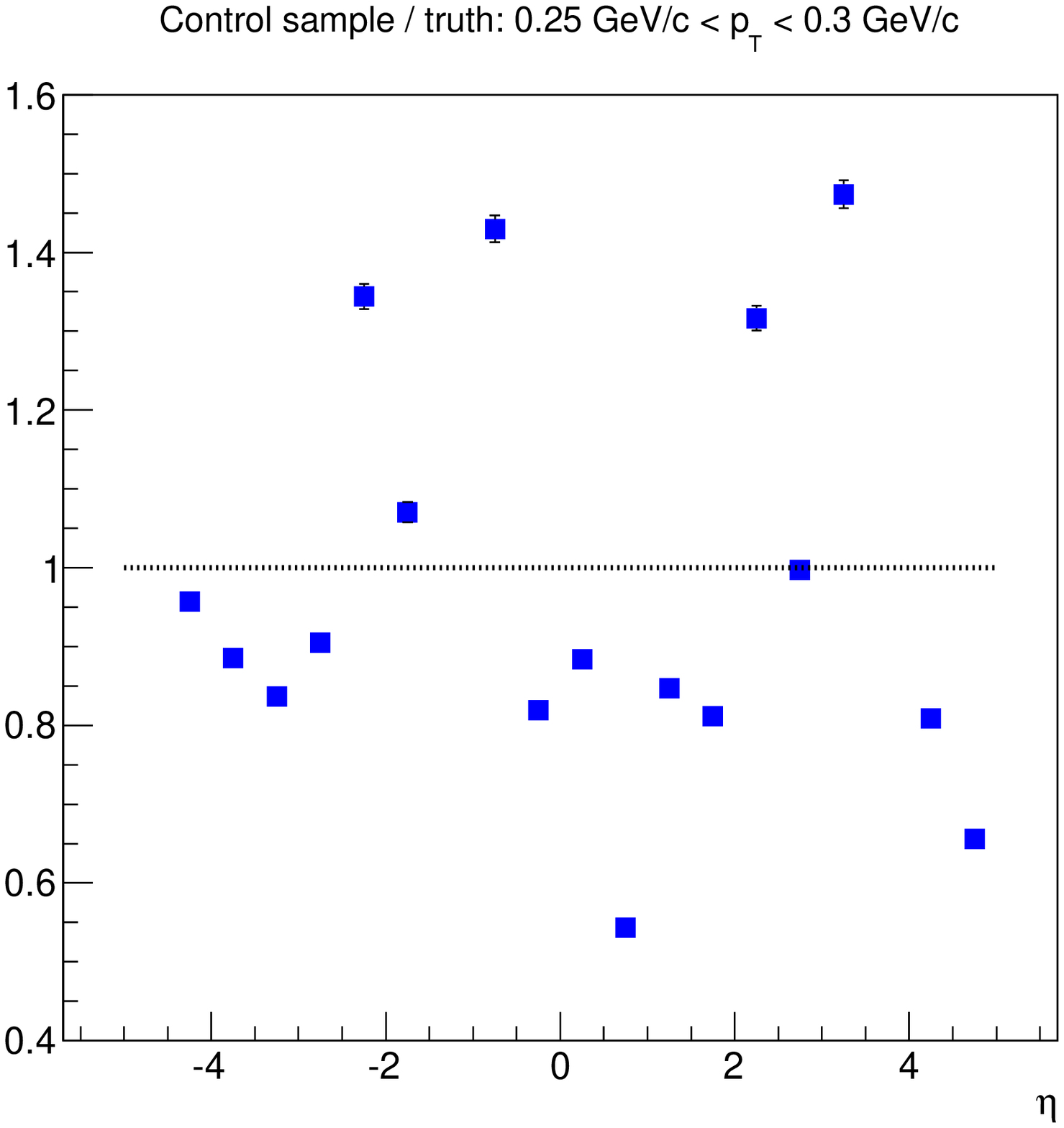}
}
\subfloat[]{
\includegraphics[scale=0.17]{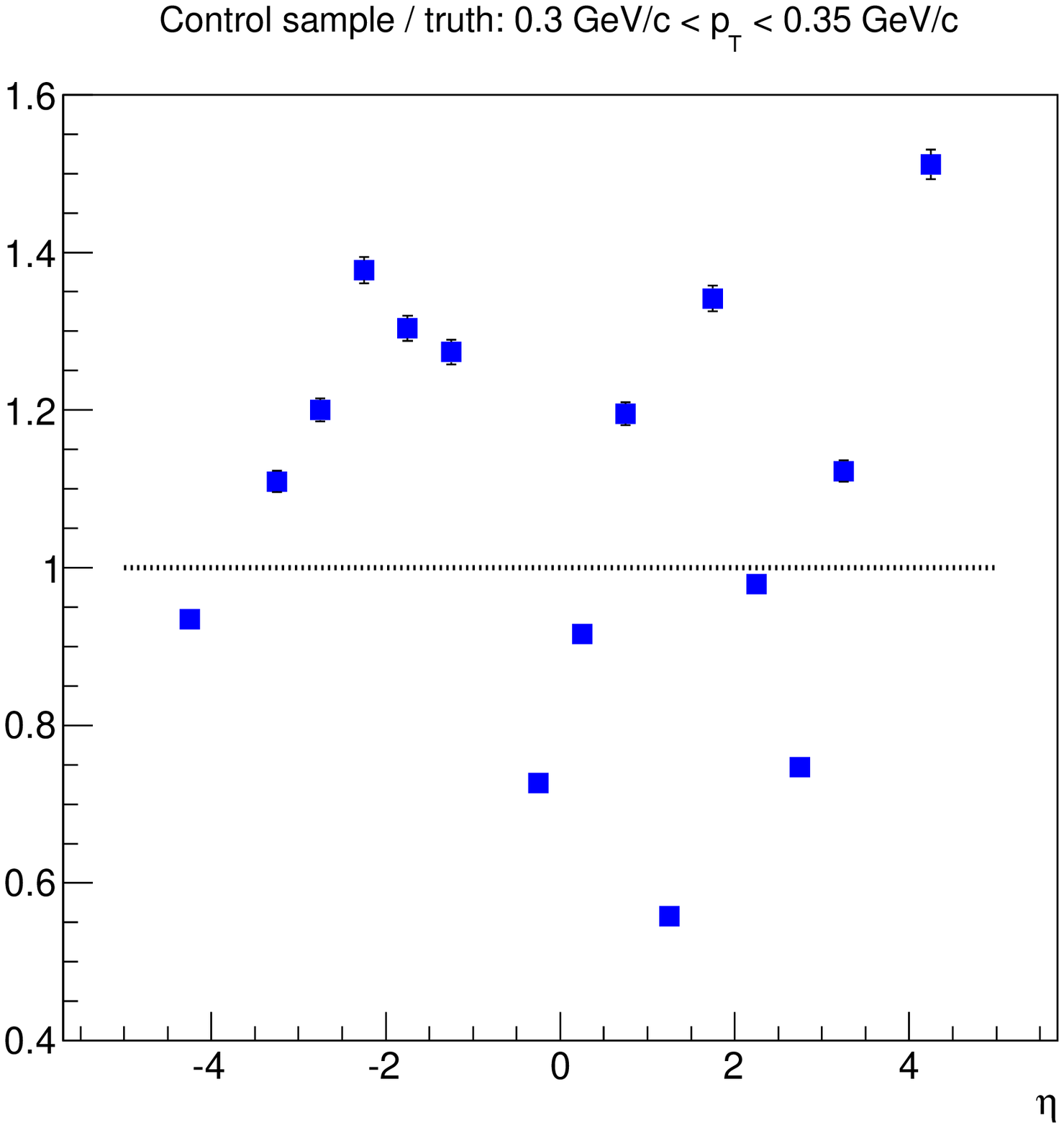}
}
\subfloat[]{
\includegraphics[scale=0.17]{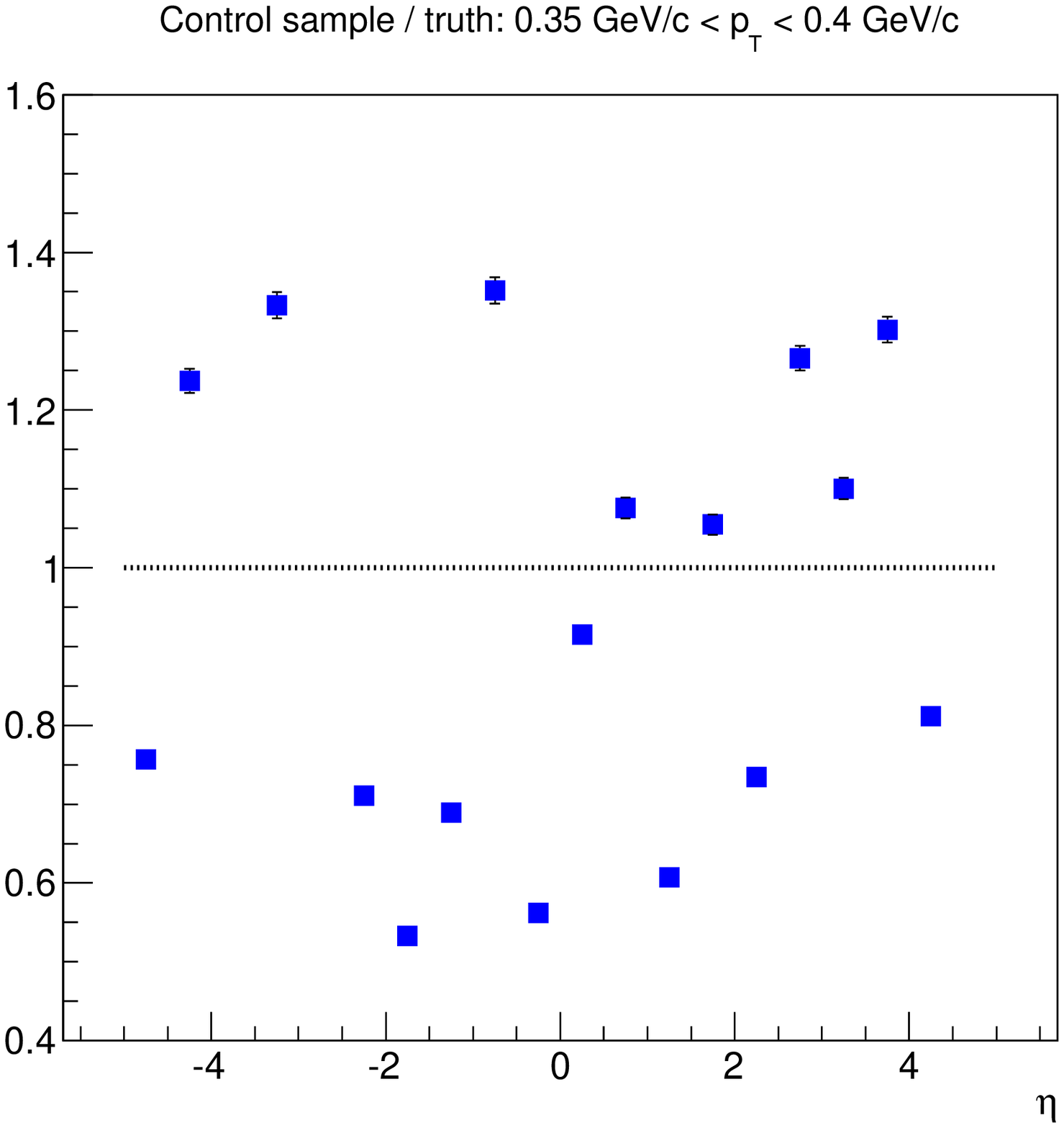}
}\\
\subfloat[]{
\includegraphics[scale=0.17]{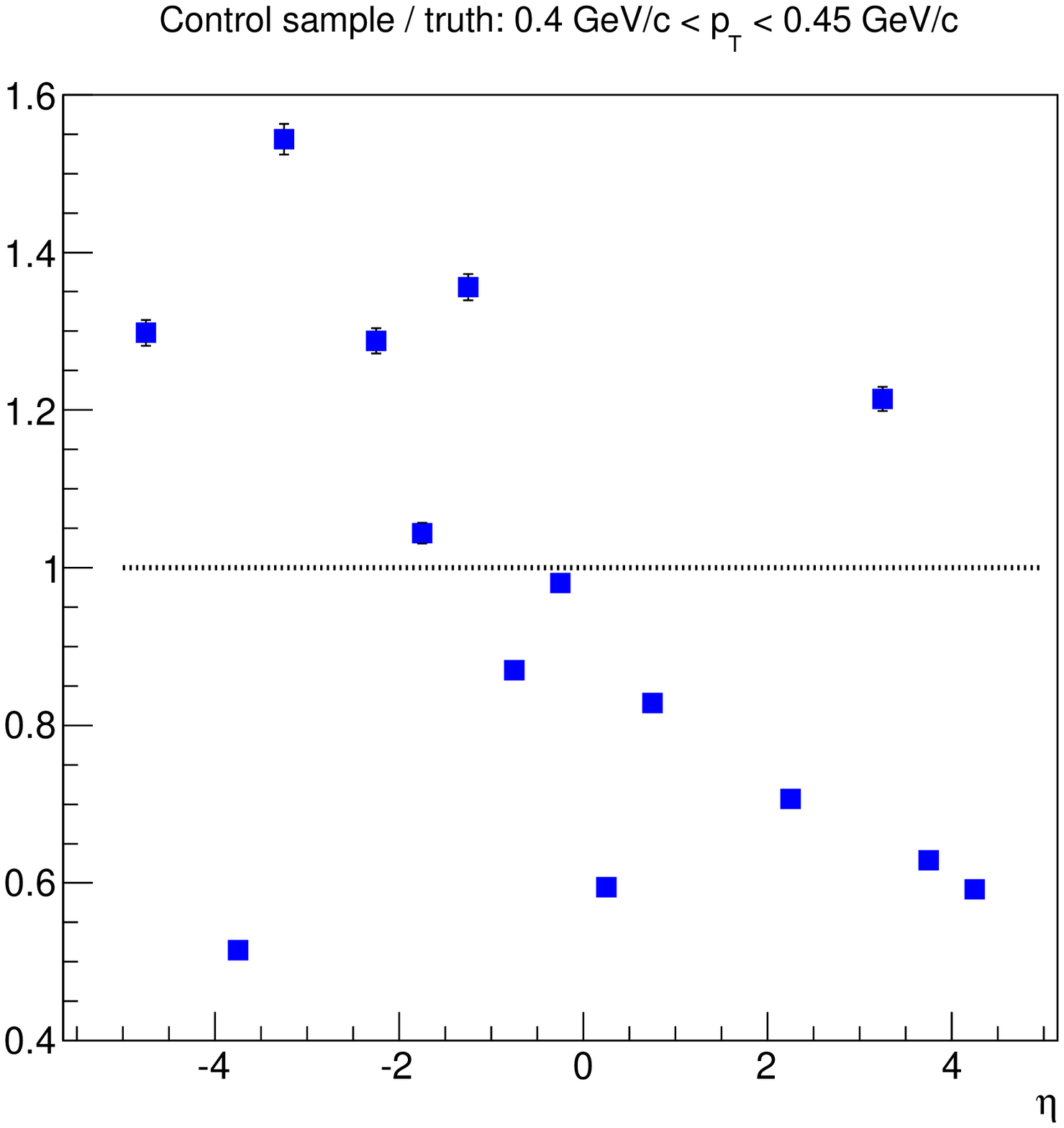}
}
\subfloat[]{
\includegraphics[scale=0.17]{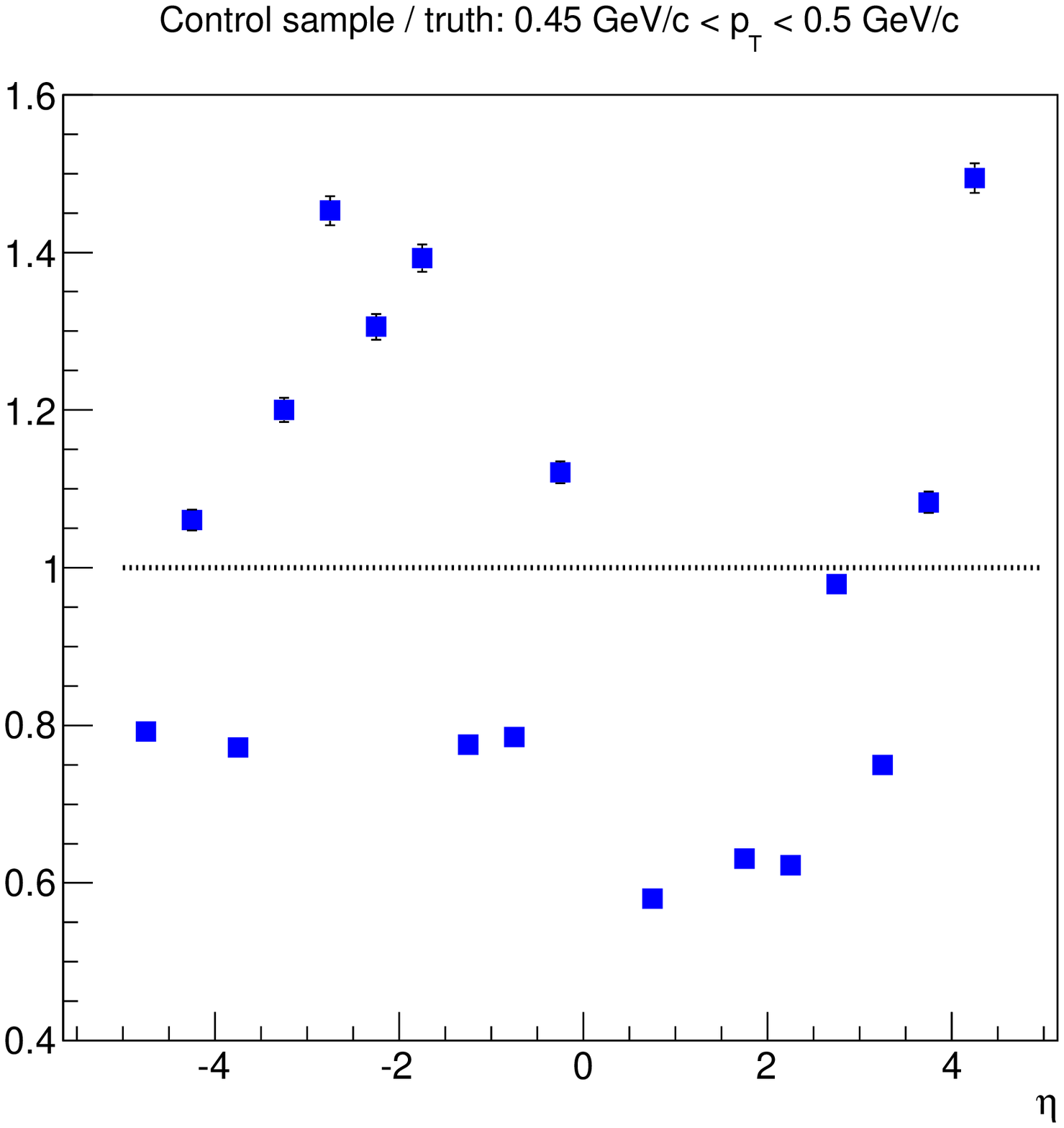}
}
\caption[]{
Ratio between the control sample and the true $(\eta, p_T)$ distribution of neutral soft QCD particles as a function of particle $\eta$. The ratios are shown in different $p_T$ bins in the region $0<p_T<0.5$~GeV/c. The distributions are normalised to unit volume. The error bars correspond to one Poisson standard deviation on the control sample bin contents as described in the text. (a) $0<p_T<0.05$~GeV/c. (b) $0.05~\mbox{GeV/c}<p_T<0.1$~GeV/c. (c) $0.1~\mbox{GeV/c}<p_T<0.15$~GeV/c. (d) $0.15~\mbox{GeV/c}<p_T<0.2$~GeV/c. (e) $0.2~\mbox{GeV/c}<p_T<0.25$~GeV/c. (f) $0.25~\mbox{GeV/c}<p_T<0.3$~GeV/c. (g) $0.3~\mbox{GeV/c}<p_T<0.35$~GeV/c. (h) $0.35~\mbox{GeV/c}<p_T<0.4$~GeV/c. (i) $0.4~\mbox{GeV/c}<p_T<0.45$~GeV/c. (j) $0.45~\mbox{GeV/c}<p_T<0.5$~GeV/c.
}
\label{fig:rats_cs}
\end{figure*}

In order to verify the agreement between the shapes of the estimated and of the true $(\eta, p_T)$ distributions of neutral soft QCD particles, we compared the estimated distribution to the true one using Monte Carlo truth information in different $p_T$ bins. Figure \ref{fig:rats_cs} displays the ratio between the control sample $(\eta, p_T)$ distribution of neutral soft QCD particles and the corresponding true distribution in the reference event as a function of particle $\eta$ in $p_T$ bins of width 0.05~GeV/c between 0 and 0.5~GeV/c. The error bars correspond to one Poisson standard deviation on the number of particles in the control sample. The plots highlight the effect of statistical fluctuations in the data, which are responsible for the observed discrepancies between the shapes of the particle-level distributions inside individual events and the ``average'' shape that corresponds to the high-statistics control sample.

\begin{figure*}
\centering
\subfloat[]{
\includegraphics[scale=0.17]{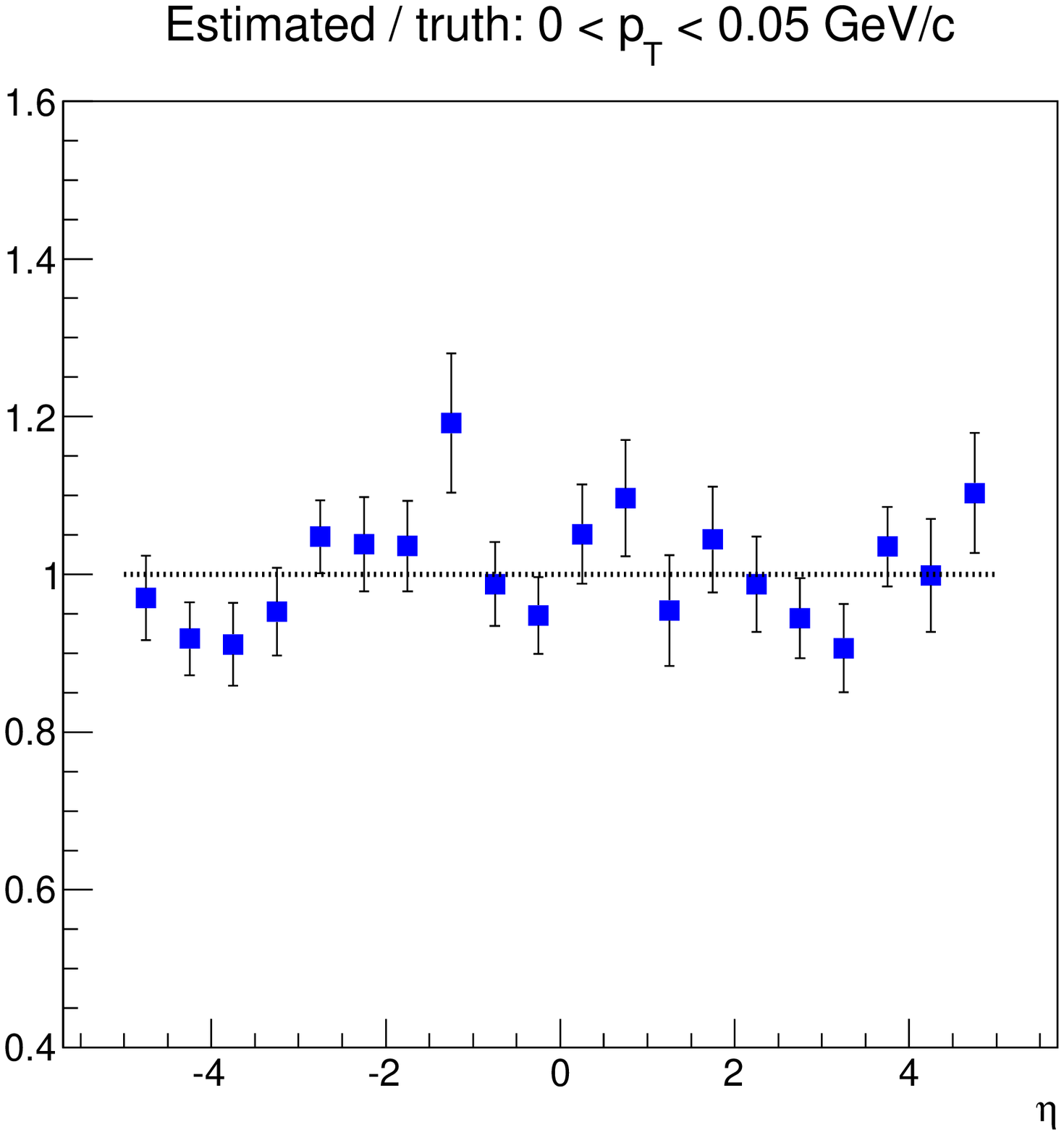}
}
\subfloat[]{
\includegraphics[scale=0.17]{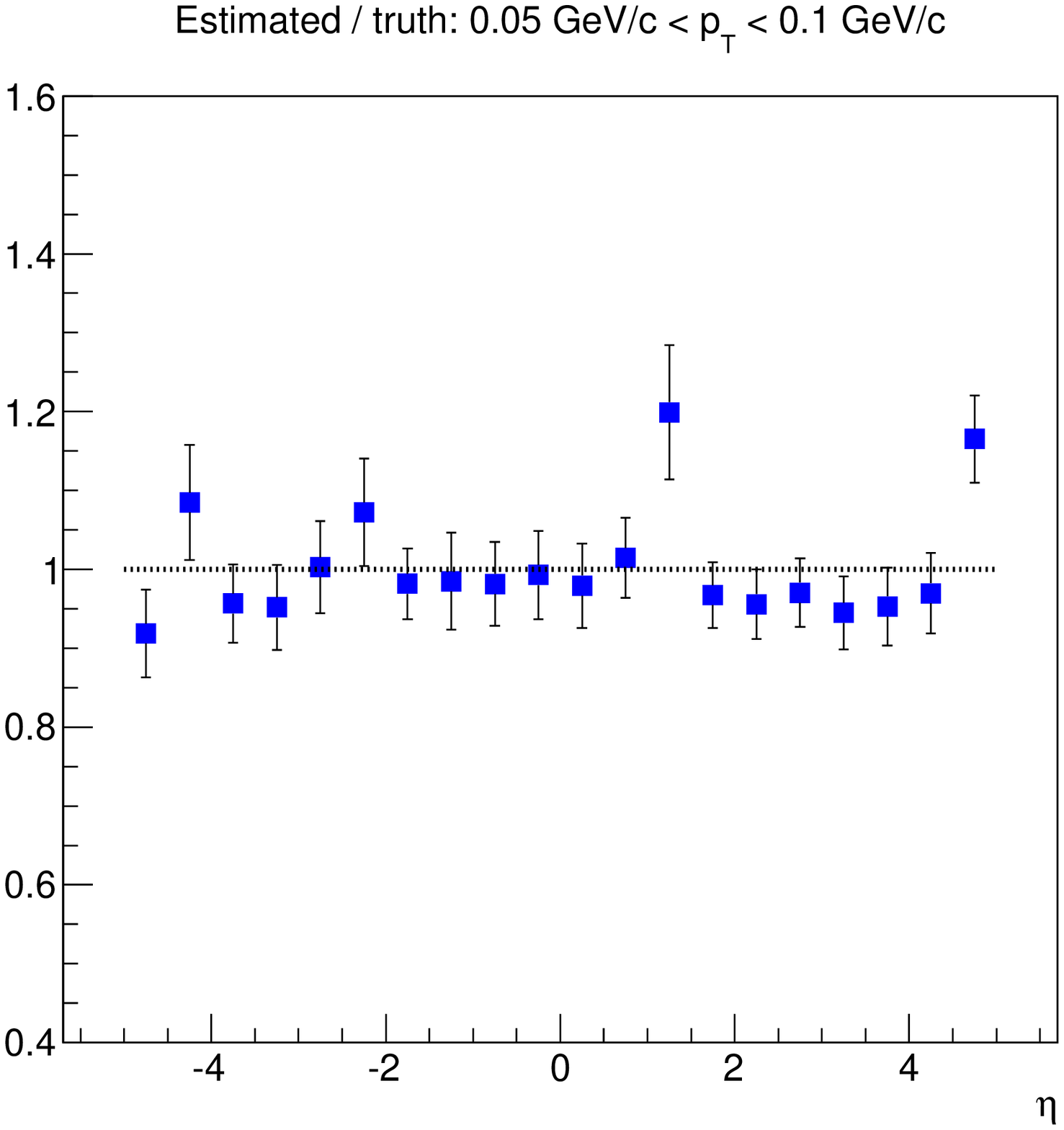}
}
\subfloat[]{
\includegraphics[scale=0.17]{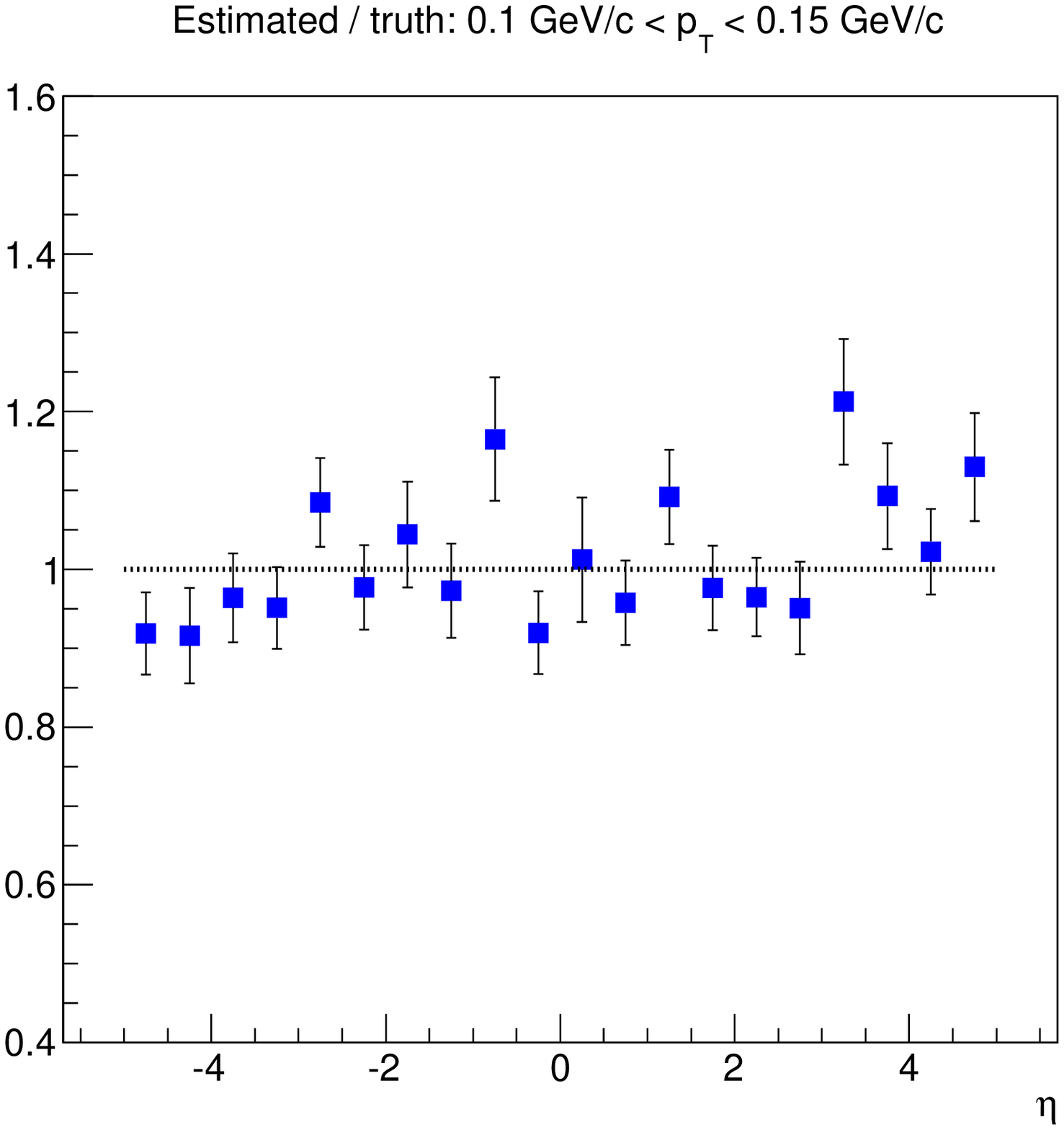}
}
\subfloat[]{
\includegraphics[scale=0.17]{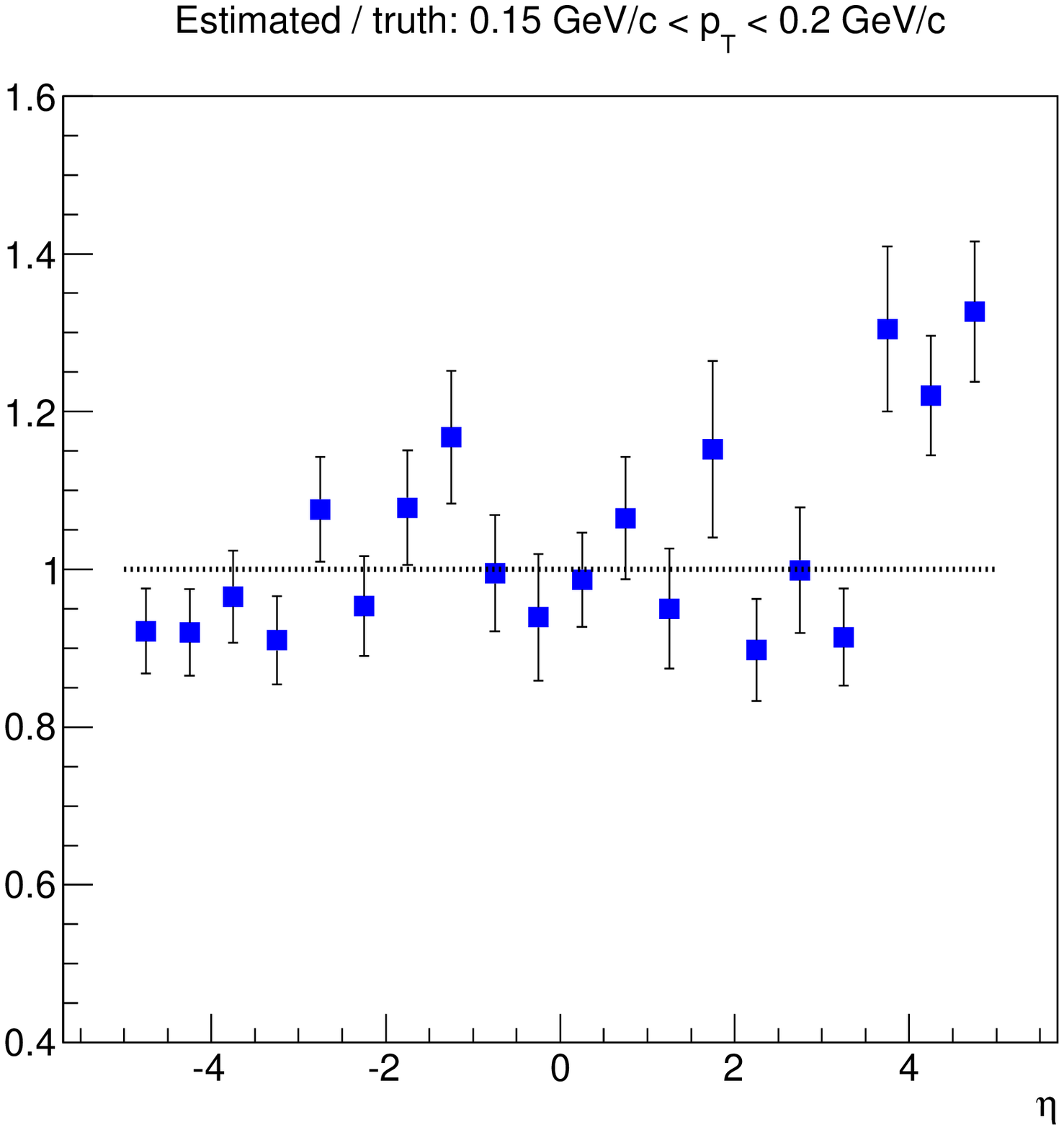}
}\\
\subfloat[]{
\includegraphics[scale=0.17]{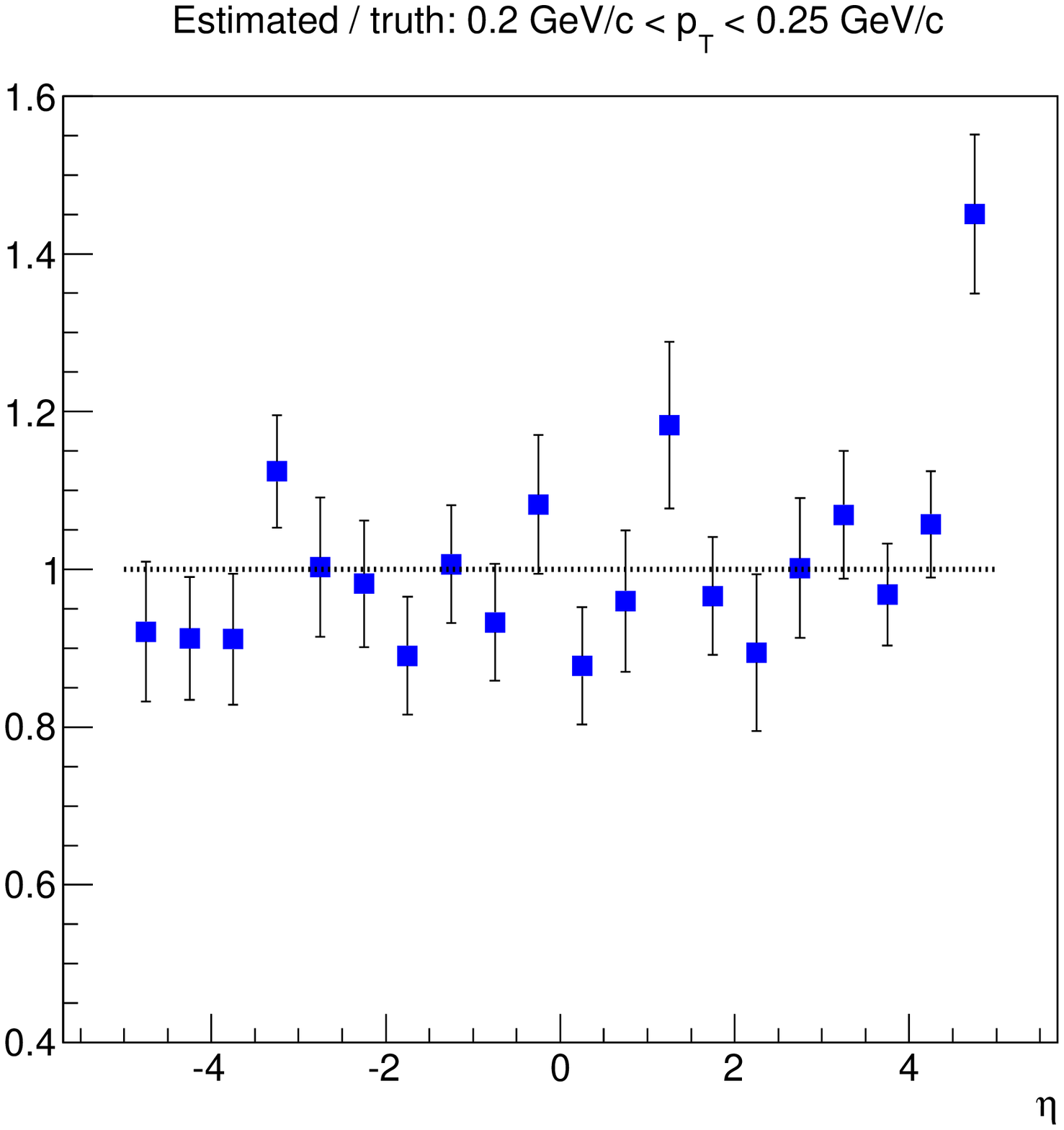}
}
\subfloat[]{
\includegraphics[scale=0.17]{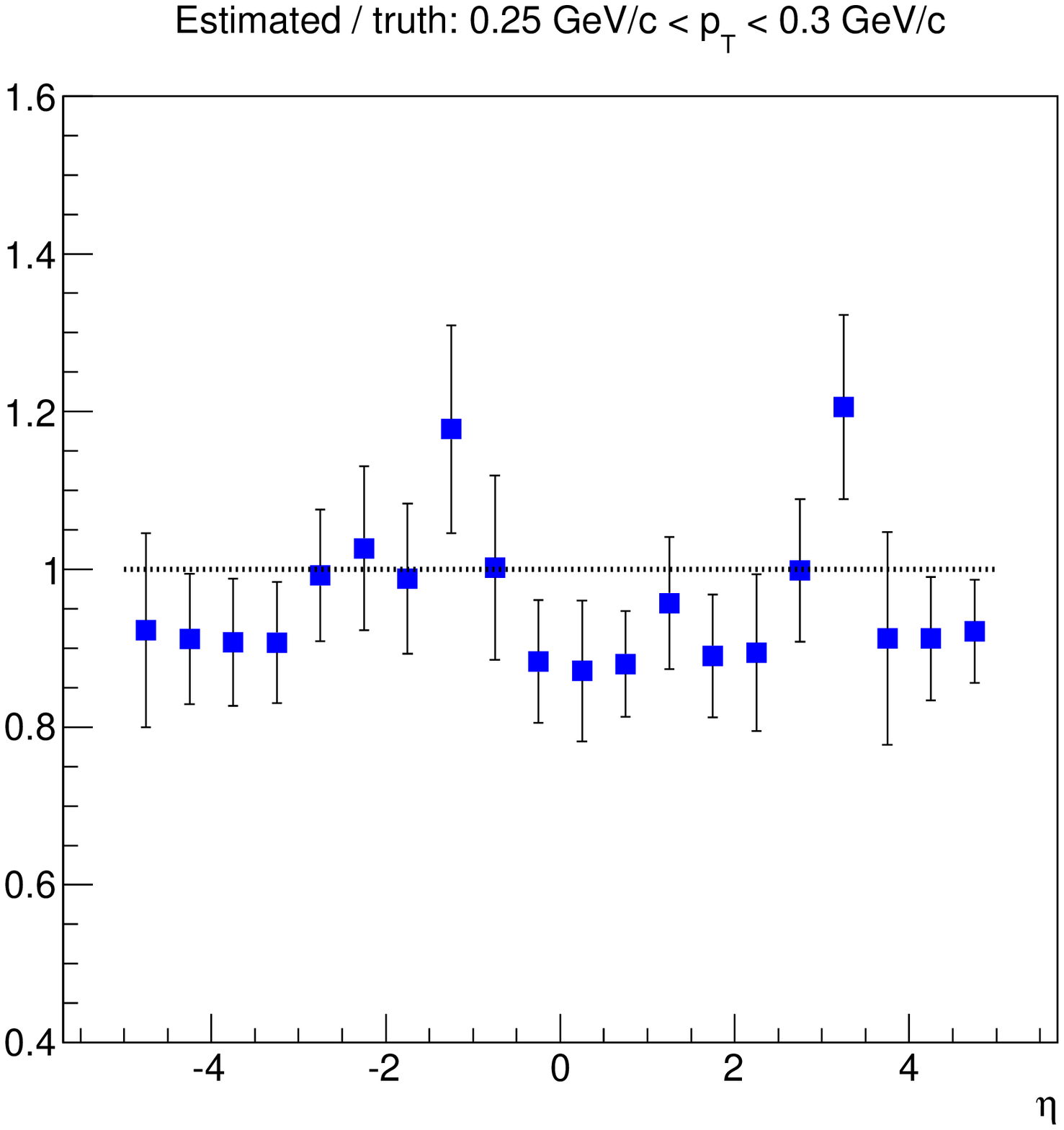}
}
\subfloat[]{
\includegraphics[scale=0.17]{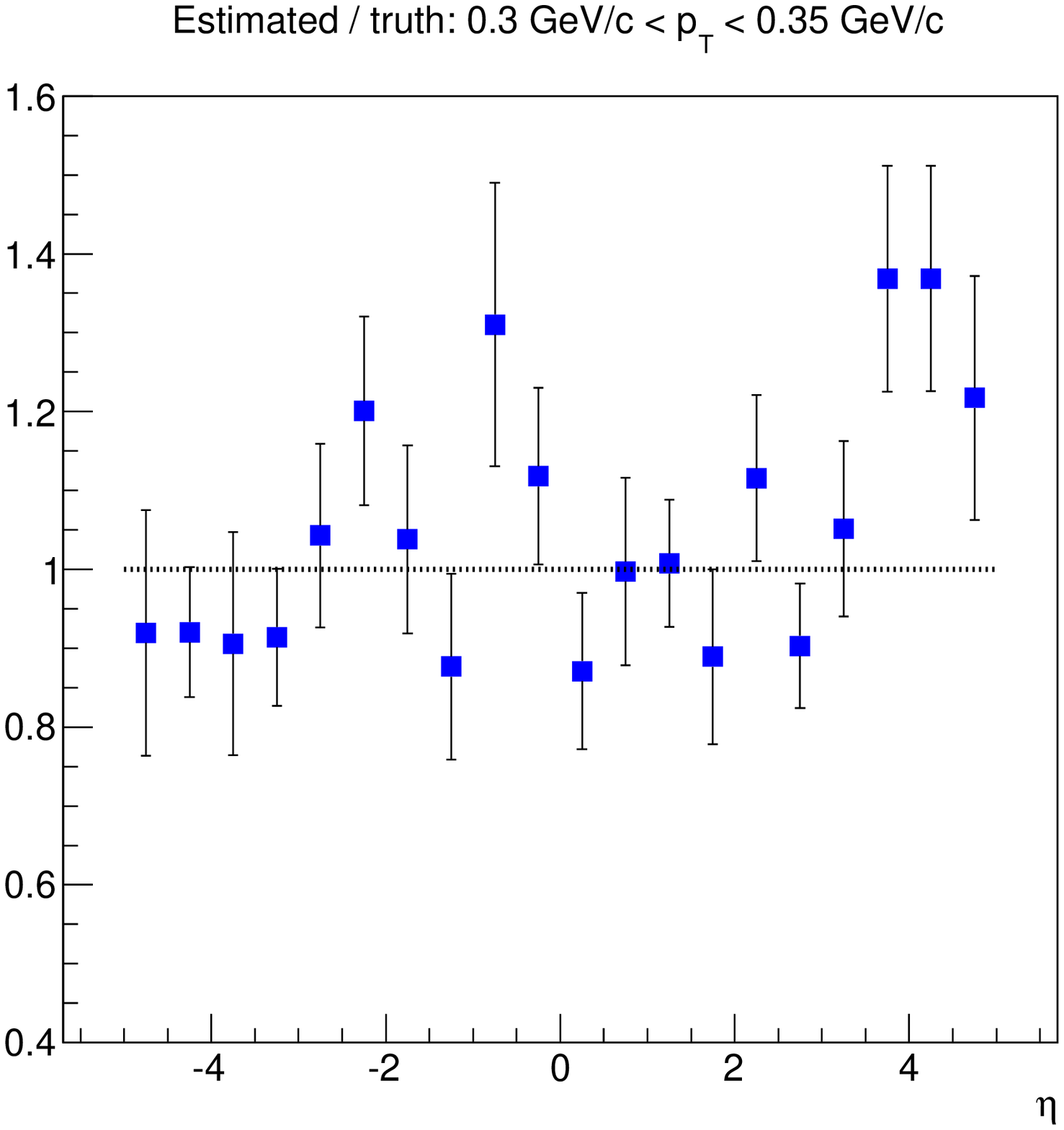}
}
\subfloat[]{
\includegraphics[scale=0.17]{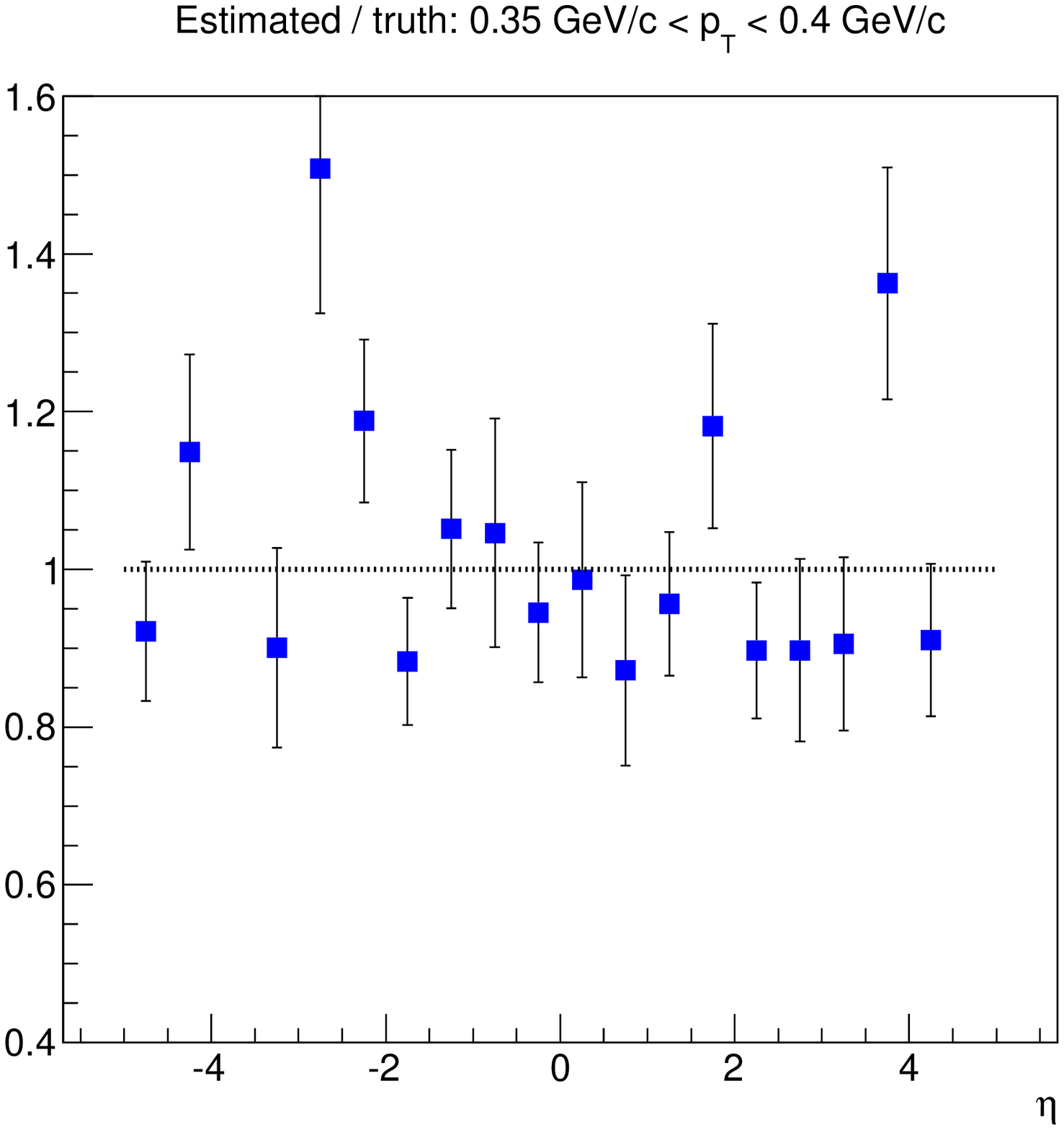}
}\\
\subfloat[]{
\includegraphics[scale=0.17]{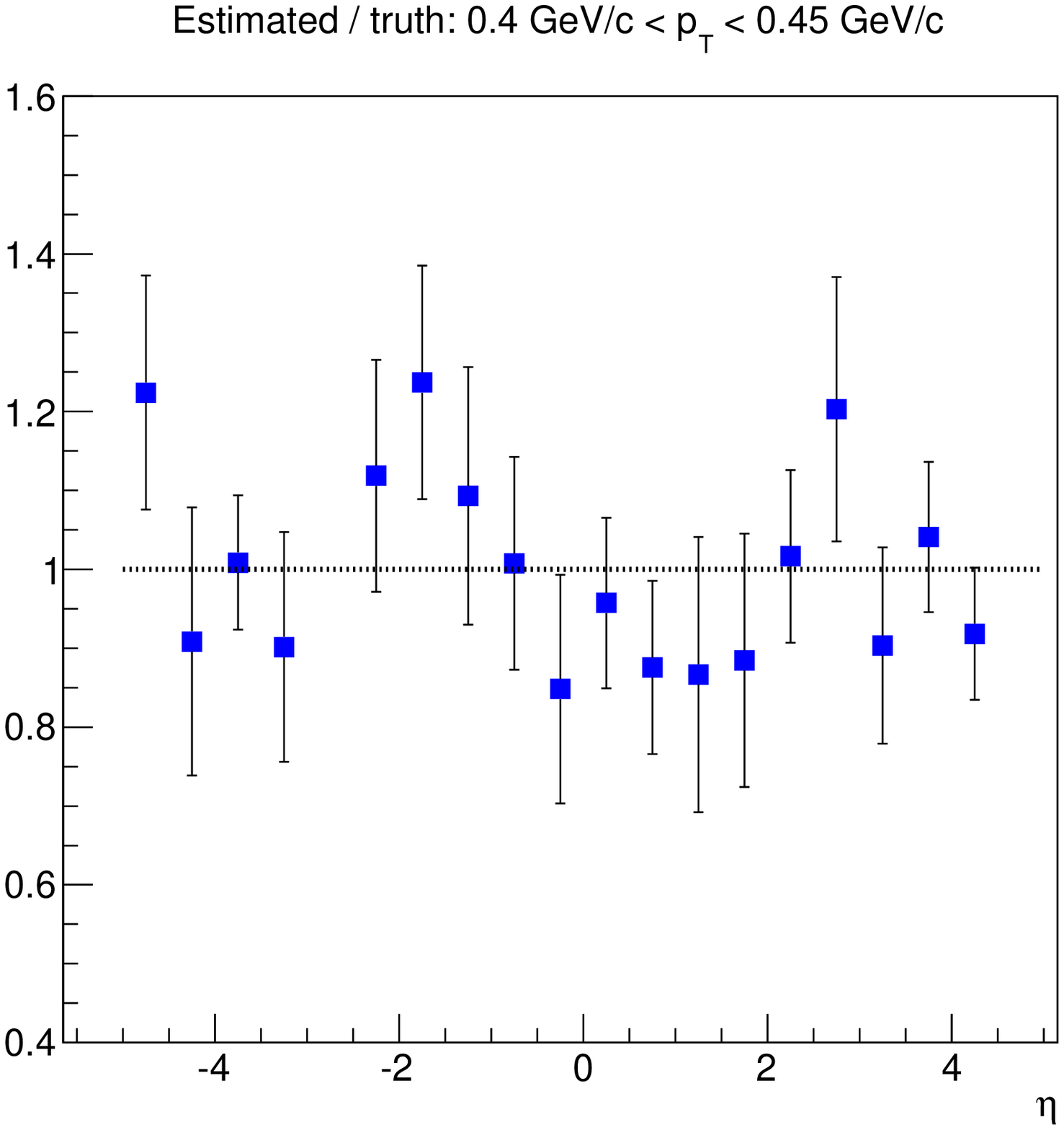}
}
\subfloat[]{
\includegraphics[scale=0.17]{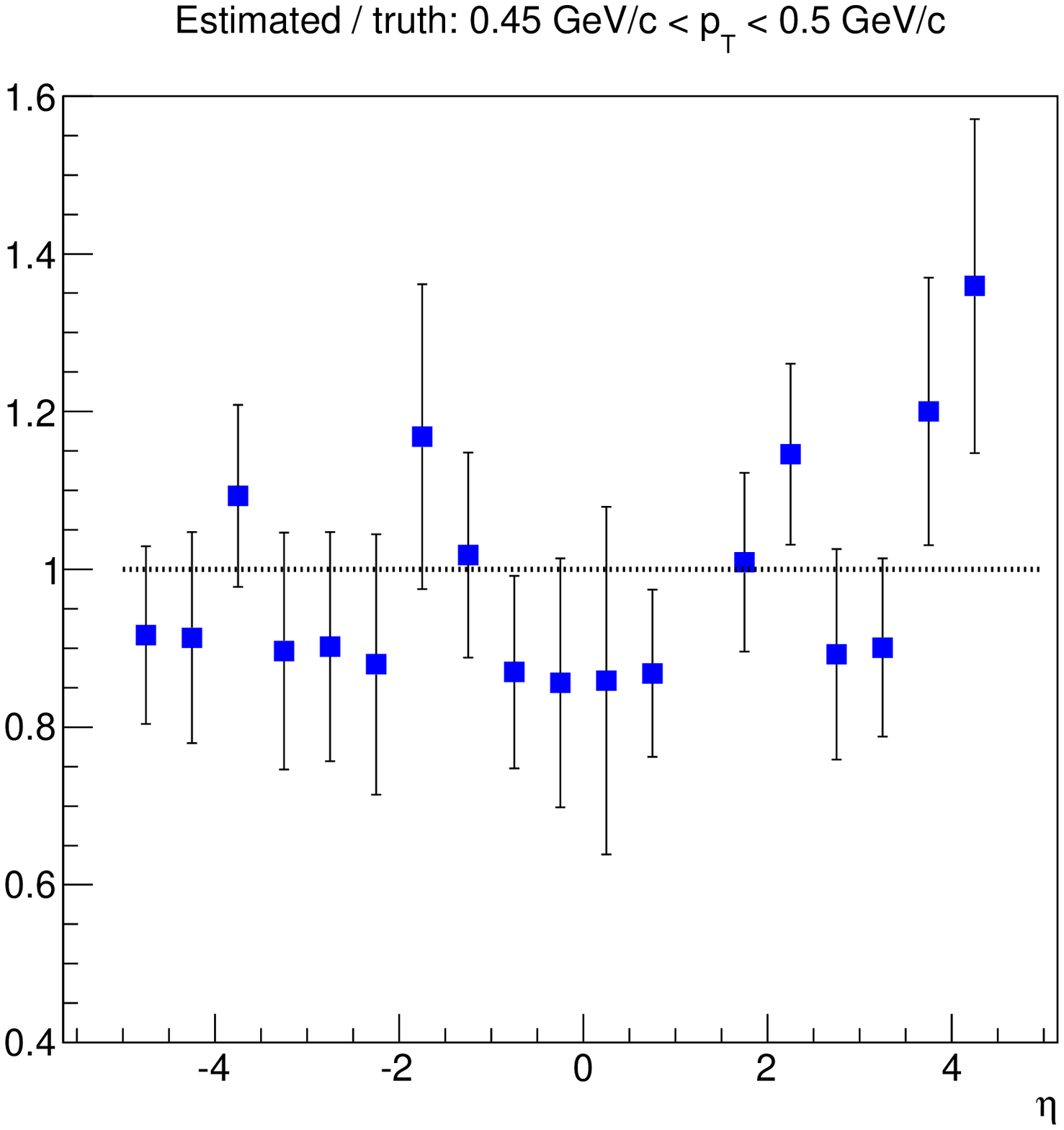}
}
\caption[]{
Ratio between the estimated $(\eta, p_T)$ distribution of neutral soft QCD particles and the corresponding true distribution. The ratios are shown in different $p_T$ bins of width 0.05~GeV/c in the region $0 < p_T < 0.5$~GeV/c. The distributions are normalised to unit volume. The error bars correspond to one binomial standard deviation on the bin contents as described in the text. (a) $0<p_T<0.05$~GeV/c. (b) $0.05~\mbox{GeV/c}<p_T<0.1$~GeV/c. (c) $0.1~\mbox{GeV/c}<p_T<0.15$~GeV/c. (d) $0.15~\mbox{GeV/c}<p_T<0.2$~GeV/c. (e) $0.2~\mbox{GeV/c}<p_T<0.25$~GeV/c. (f) $0.25~\mbox{GeV/c}<p_T<0.3$~GeV/c. (g) $0.3~\mbox{GeV/c}<p_T<0.35$~GeV/c. (h) $0.35~\mbox{GeV/c}<p_T<0.4$~GeV/c. (i) $0.4~\mbox{GeV/c}<p_T<0.45$~GeV/c. (j) $0.45~\mbox{GeV/c}<p_T<0.5$~GeV/c.
}
\label{fig:rats_sculpt}
\end{figure*}

The corresponding ratio between the estimated and the true distribution in the reference event is displayed in figure \ref{fig:rats_sculpt}, which shows a significantly-improved agreement. The error bars are calculated based on (\ref{eq:dnsculpt}). 

\begin{center}
\begin{table*}
\caption{
Subset of the results obtained on the event chosen to illustrate the performance of this technique. The figures correspond to $(\eta, p_T)$ bins with $0 < p_T < 0.15$~GeV/c and $\left|\eta\right|\leq 2.5$, and with at least two particles in the data, i.e. $n(\eta, p_T)\geq 2$. The columns correspond to the centres of the $\eta$ and $p_T$ bins, to $n_s^*$, $n_b^*$, $w_0$, $\hat{n}_b$, $\sigma_{\hat{n}_b}$, $\Delta w$, and $p_T\Delta w$.
}
\hfill{}
\begin{tabular}{c c c c c c c c c}
\hline
$\eta$ & $p_T$ (GeV/c) & $n_s^*$ & $n_b^*$ & $w_0$ & $\hat{n}_b$ & $\sigma_{\hat{n}_b}$ & $\Delta w$ & $p_T\Delta w$ (GeV/c)\\
\hline
-2.25 & 0.025 & 3 & 20 & 0.946 & 21.7 & 1.1 & 0.022 & 0.0005\\
-2.25 & 0.075 & 3 & 16 & 0.946 & 18 & 1 & 0.039 & 0.0029\\
-2.25 & 0.125 & 2 & 24 & 0.945 & 24.6 & 1.2 & -0.011 & -0.0014\\
-1.75 & 0.025 & 4 & 25 & 0.938 & 27.2 & 1.3 & 0.019 & 0.0005\\
-1.75 & 0.075 & 3 & 34 & 0.945 & 35 & 1.4 & -0.009 & -0.0007\\
-1.75 & 0.125 & 3 & 18 & 0.939 & 19.7 & 1.1 & 0.023 & 0.0029\\
-1.25 & 0.025 & 4 & 12 & 0.938 & 15 & 1 & 0.093 & 0.0023\\
-1.25 & 0.075 & 2 & 20 & 0.94 & 20.7 & 1.1 & -0.008 & -0.0006\\
-1.25 & 0.125 & 2 & 22 & 0.937 & 22.5 & 1.2 & -0.016 & -0.002\\
-0.75 & 0.025 & 3 & 28 & 0.937 & 29 & 1.4 & -0.007 & -0.0002\\
-0.75 & 0.075 & 3 & 29 & 0.935 & 29.9 & 1.4 & -0.011 & -0.0008\\
-0.75 & 0.125 & 5 & 16 & 0.933 & 19.6 & 1.1 & 0.08 & 0.01\\
-0.25 & 0.025 & 2 & 32 & 0.937 & 31.9 & 1.4 & -0.03 & -0.0007\\
-0.25 & 0.075 & 3 & 26 & 0.935 & 27.1 & 1.3 & -0.005 & -0.0004\\
-0.25 & 0.125 & 1 & 28 & 0.934 & 27.1 & 1.3 & -0.047 & -0.0059\\
0.25 & 0.025 & 4 & 22 & 0.935 & 24.3 & 1.3 & 0.026 & 0.0006\\
0.25 & 0.075 & 3 & 29 & 0.933 & 29.9 & 1.4 & -0.013 & -0.001\\
0.25 & 0.125 & 2 & 14 & 0.932 & 14.9 & 1 & 0.005 & 0.0006\\
0.75 & 0.025 & 4 & 17 & 0.933 & 19.6 & 1.1 & 0.048 & 0.0012\\
0.75 & 0.075 & 4 & 30 & 0.94 & 31.9 & 1.4 & 0.008 & 0.0006\\
0.75 & 0.125 & 2 & 27 & 0.936 & 27.2 & 1.3 & -0.024 & -0.0031\\
1.25 & 0.025 & 1 & 15 & 0.939 & 15 & 1 & -0.026 & -0.0006\\
1.25 & 0.075 & 4 & 12 & 0.943 & 15.1 & 0.9 & 0.098 & 0.0073\\
1.25 & 0.125 & 5 & 23 & 0.941 & 26.3 & 1.2 & 0.048 & 0.006\\
1.75 & 0.025 & 3 & 18 & 0.939 & 19.7 & 1.1 & 0.023 & 0.0006\\
1.75 & 0.075 & 3 & 40 & 0.943 & 40.6 & 1.5 & -0.017 & -0.0013\\
1.75 & 0.125 & 2 & 24 & 0.944 & 24.6 & 1.2 & -0.012 & -0.0015\\
2.25 & 0.025 & 2 & 20 & 0.941 & 20.7 & 1.1 & -0.006 & -0.0002\\
2.25 & 0.075 & 2 & 34 & 0.945 & 34 & 1.4 & -0.023 & -0.0017\\
2.25 & 0.125 & 2 & 28 & 0.944 & 28.3 & 1.3 & -0.019 & -0.0023\\
\hline
\end{tabular}
\hfill{}
\label{tab:stats}
\end{table*}
\end{center}

%
%
\vskip -1.2cm

Table \ref{tab:stats} further illustrates the performance of this technique on the same event presented in the plots. The figures in the table correspond to bins with $0 < p_T < 0.15$~GeV/c and $\left|\eta\right|\leq 2.5$ with at least two particles in the data, i.e. $n(\eta, p_T)\geq 2$. The columns correspond to the centres of the $\eta$ and $p_T$ bins, to $n_s^*$, $n_b^*$, $w_0$, $\hat{n}_b$, $\sigma_{\hat{n}_b}$, $\Delta w = w_0-w^*$, and to $p_T\Delta w$.

As pointed out in section \ref{weights}, although this approach is being presented with reference to scenarios where the number of signal particles is on average much lower than the number of soft QCD particles, the task of estimating the latter becomes trivial in the limit where the number of signal particles is zero, since in that case $w_0(\eta, p_T)=1$ and $\hat{n}_b(\eta, p_T)=n^*(\eta, p_T)$. However, in practice, it is not known which bins contain particles originating from the signal hard scattering and which do not.

\begin{figure*}
\centering
\begin{minipage}{17pc}
\includegraphics[scale=0.37]{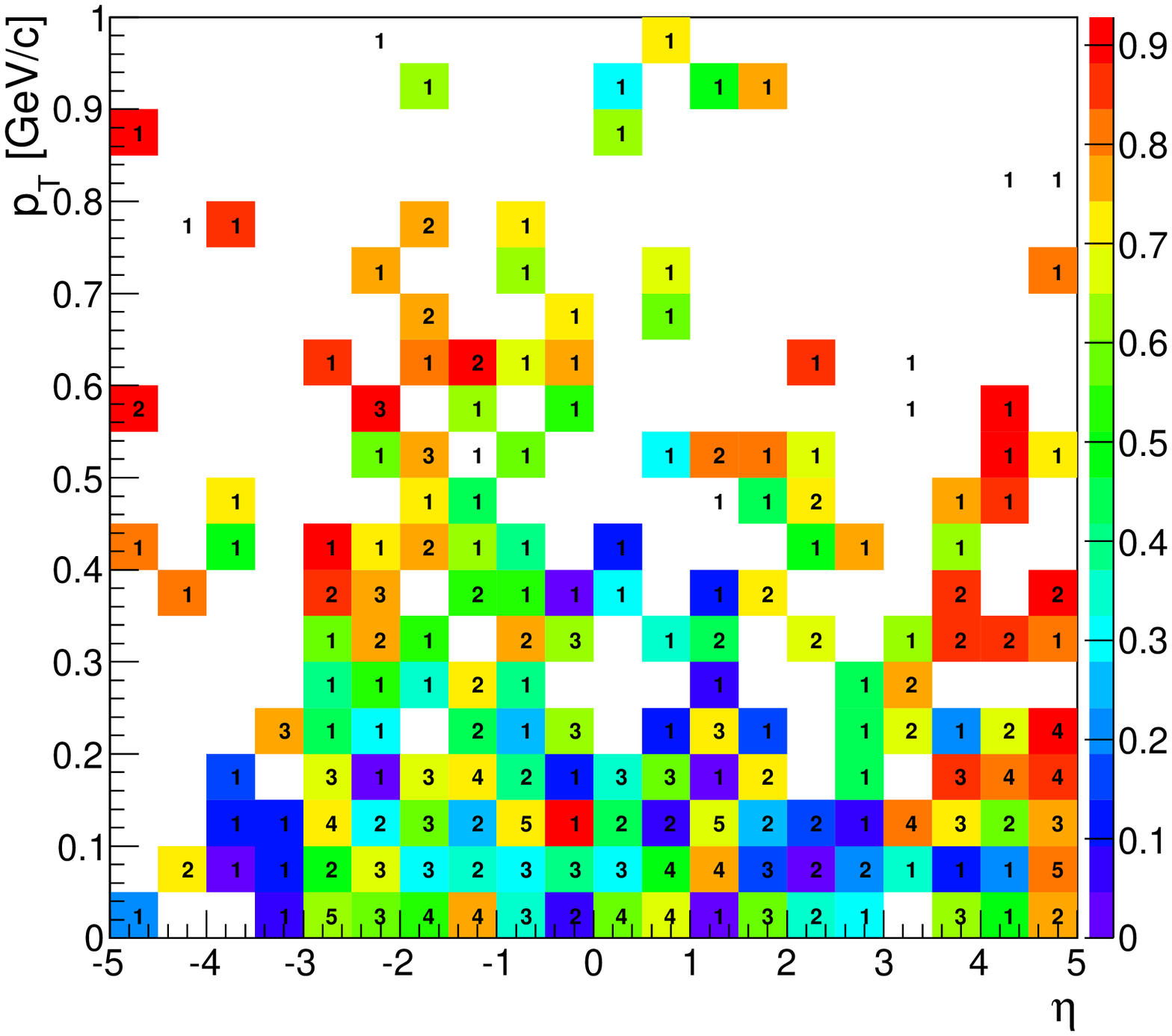}
\end{minipage}\hspace{3pc}%
\begin{minipage}{14pc}
\caption{\label{fig:delta_nb_over_ns}True number of particles originating from the signal hard scattering, $n_s^*(\eta, p_T)$, superimposed with a heatmap corresponding to $\left|\Delta \hat{n}_b\right| / n_s^*$ in those bins that contain more than one particle in the data, at least one of which originating from the signal hard scattering. Additional information is given in the text.}
\end{minipage}
\end{figure*}

With a view to verifying that our results are more accurate than those that would be obtained if the presence of signal particles in the data was neglected, the deviation of the estimated number of neutral soft QCD particles from the corresponding true value, $\Delta \hat{n}_b = \hat{n}_b - n^*_b$, was compared to the true number of signal particles, $n_s^*$. Figure \ref{fig:delta_nb_over_ns} displays $\left|\Delta \hat{n}_b\right| / n_s^*$ in those bins that contain more than 1 particle in the data, at least one of which originating from the signal hard scattering. The absolute difference between the estimated number of neutral soft QCD particles and the unknown true number averaged over the events analysed in this study was found to be $\left<\left|\Delta\hat{n}_b\right|\right>_{avg}\simeq0.6$, where $\left<\cdot\right>$ denotes the average over those bins that contain at least 1 background particle, and the subscript ``$avg$'' refers to the average over the events. The average absolute error on $\hat{n}_b$ on the data set analysed was therefore found to be lower than 1 particle.

\begin{figure*}
\centering
\subfloat[]{
\includegraphics[scale=0.37]{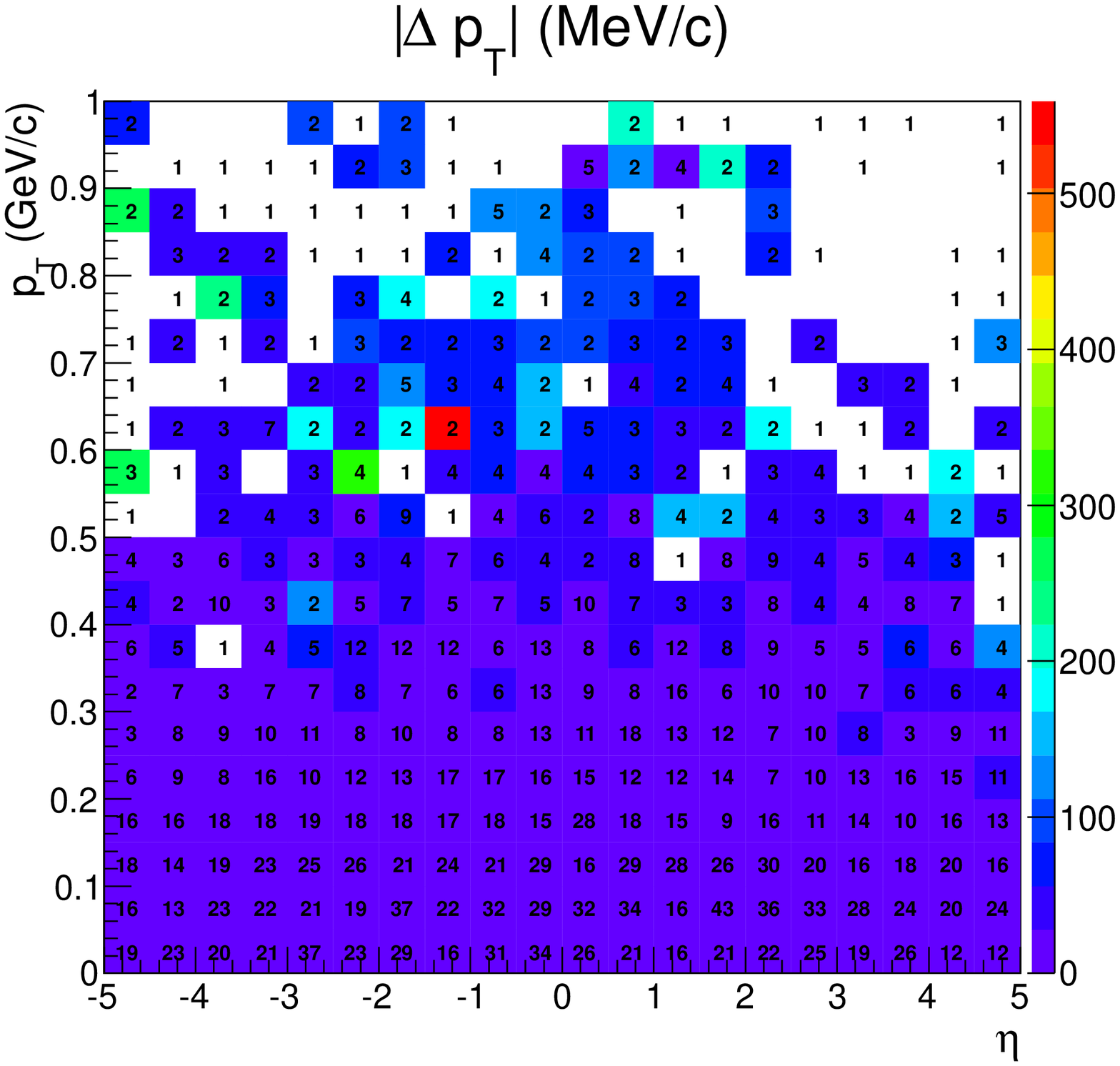}
}
\subfloat[]{
\includegraphics[scale=0.37]{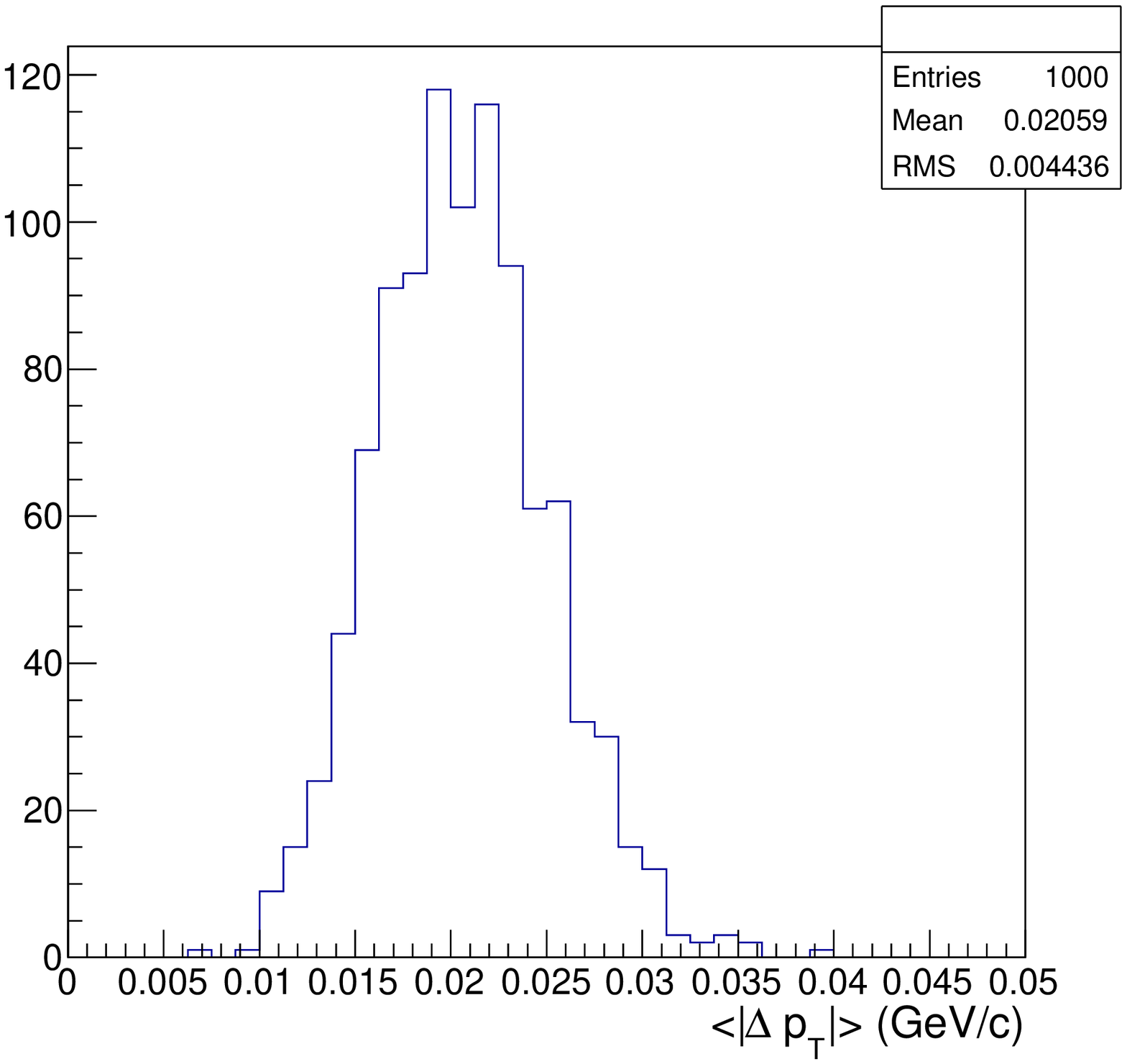}
}
\caption[]{
(a) Absolute difference between the estimated and the true particle weight multiplied by the particle $p_T$, $\left|\Delta p_T\right|$, shown as a heatmap superimposed to the estimated number of neutral soft QCD particles, $\hat{n}_b(\eta, p_T)$. As expected, the rescaled particle $p_T$, $w_0 p_T$, is a better estimate of the true value in the low-$p_T$ region, which is more densely populated by soft QCD particles. (b) Distribution of the average $\left|\Delta p_T\right|$, $\left<\left|\Delta p_T\right|\right>$, over the events analysed in this study. The average is weighted with the number of particles in the data, $n(\eta, p_T)$, and is taken over the $(\eta, p_T)$ space. Additional information is provided in the text.
}
\label{fig:delta_pT}
\end{figure*}

It should also be emphasised that, although we are not explicitly proposing our algorithm with reference to pileup subtraction inside jets, we also envisage the possibility of combining this method with state-of-the-art jet calibration techniques, e.g. using local estimates of neutral pileup particle multiplicity as constraints in jet substructure algorithms.

\subsection{Missing transverse energy resolution}

In this section, we employ a similar approach to \cite{PUPPI} whereby the weights are used to rescale the particle four-momentum vectors, with a view to assessing the impact of the weights defined in section \ref{weights} on the resolution of the missing transverse energy, $\slashed{E}_T$.

A full analysis of the impact on $\slashed{E}_T$ resolution at the LHC is outside the scope of this article. We are here providing a preliminary estimate concentrating on the effect of pileup, assuming 50 vertices per event.

It is worth recalling that the measurement of $\slashed{E}_T$ relies on information that is provided by independent sources, such as the calorimeters, the trackers and the muon subdetectors, and that it is sensitive both to pileup contamination and to beam-induced effects \cite{ETmiss}.

In particular, $\slashed{E}_T$ is one of the observables that are most significantly affected by contamination from soft QCD interactions at hadron colliders. It is estimated that each additional pileup interaction at the LHC adds $3.3\div 3.7$~GeV in quadrature to the Particle Flow $\slashed{E}_T$ resolution and, although existing methods have been shown to be extremely useful below 35 vertices per event \cite{REVIEW_PILEUP_2014}, there is still a margin for improvement. Moreover, it is not clear what the performance of the existing techniques is going to be like as the number of vertices per event increases.

The pileup-related contribution to the $\slashed{E}_T$ resolution associated with the use of the weights defined in section \ref{weights} is a function of the deviation of $w_0(\eta, p_T)$ from the corresponding true value, $w^*(\eta, p_T)$. Specifically, when $w_0(\eta, p_T)$ is used to rescale the four-momentum of a particle with transverse momentum $p_T$, a deviation of the weight from its true value results in a fraction of the particle $p_T$ being assigned to the wrong physics process. If $\Delta w(\eta, p_T) = w_0(\eta, p_T)-w^*(\eta, p_T)$ denotes the deviation of the particle weight from its true value, the amount of incorrectly-assigned $p_T$ for that particle is given by $\Delta p_T(\eta, p_T) = p_T\Delta w(\eta, p_T)$.

Figure \ref{fig:delta_pT} (a) displays the estimated number of neutral soft QCD particles across the $(\eta, p_T)$ space, superimposed with a heatmap corresponding to $\Delta p_T(\eta, p_T)$ in MeV/c. A comparison with figure \ref{fig:heatmap} (b) shows that those bins that exhibit a lower relative uncertainty on the estimated number of soft QCD particles, $\sigma_{\hat{n}_b}/\hat{n}_b$, also correspond to lower $\Delta p_T$, as expected. It is worth noticing that $\sigma_{\hat{n}_b}/\hat{n}_b = \sqrt{(1-w_0)/nw_0}$ does not depend on Monte Carlo truth information and can be estimated directly from the data. This makes it possible to restrict the use of this technique to $(\eta,p_T)$ bins that are associated with a value of $\sigma_{\hat{n}_b}/\hat{n}_b$ lower than a predefined threshold.

The quantity $\Delta p_T(\eta, p_T)$ provides an estimate of the contribution of individual particles in the event to the $\slashed{E}_T$ resolution with reference to pileup contamination. A preliminary estimate of the impact of $\Delta w$ on the $\slashed{E}_T$ resolution can be given in terms of $\sigma_{\slashed{E}_T}^{PU} \simeq \sqrt{N}<\sigma_{\Delta_{p_T}}>$, where $N$ is the total number of particles in the event and $<\sigma_{\Delta_{p_T}}>$ is the standard deviation of $\Delta p_T(\eta, p_T)$ averaged over the $(\eta, p_T)$ space. For the sake of an approximate calculation, we assume $<\sigma_{\Delta p_T}> \simeq \sigma_{<\Delta p_T>}$. 

Figure \ref{fig:delta_pT} (b) shows the distribution of $\left<\Delta p_T(\eta, p_T)\right>$ over the Monte Carlo events analysed in this study. In each event, the average in $\left<\Delta p_T(\eta, p_T)\right>$ is weighted with the number of particles in the data, $n(\eta, p_T)$, and is taken over the $(\eta, p_T)$ space. Assuming a total number of particles in the event\footnote{Both charged and neutral} $N\simeq 4,000$, the RMS of $\left<\Delta p_T(\eta, p_T)\right>$ in figure \ref{fig:delta_pT} (b) leads to $\sigma_{\slashed{E}_T}^{PU}\simeq 0.3$~GeV/c. This is a notable improvement when compared to the state of the art \cite{ETmiss}, and is in line with the results obtained using other particle weighting methods \cite{PUPPI}.

It should be emphasised that the above remarks relate to a generator-level study and exclusively concentrate on the effect of pileup. While the latter is a major source of concern in the upcoming higher-luminosity regimes at the LHC, a proper investigation will also have to take into account other factors, such as detector effects, mis-reconstrution, beam-related events, and contamination from cosmic muons. Moreover, a full investigation of this technique in a proper analysis framework will need to be performed.

\section{Conclusions and outlook\label{concl}}

We have presented a proof of concept of a new approach to the use of particle weights at high-luminosity hadron collider experiments, a distinctive feature of which is the idea of employing the weights to reshape the particle-level kinematic distributions in the data. We have applied this method to the task of estimating the number of neutral particles associated with pileup, i.e. with low-energy strong interactions from other proton-proton collisions, in different kinematic regions inside collision events. Pileup is a major source of contamination at the Large Hadron Collider, and its impact on physics analysis is expected to become even more significant in the upcoming higher-luminosity regimes. 

We build on a view of collision events as mixtures of particles originating from different physics processes, whereby the use of particle weights helps resolve the conceptual issues associated with colour connection.

Because of the quantum nature of the underlying physics processes, the kinematic patterns of pileup particles are typically different in different events, i.e. individual events can be associated with distinctive pileup kinematic ``fingerprints'' at the particle level. We have shown that our approach makes it possible to estimate the number of neutral pileup particles in different kinematic regions inside events with reasonable accuracy, regardless of whether or not particles originating from the signal hard scattering are present. Since the estimates do not correspond to average numbers of particles, but rather to the actual numbers in each event, our approach takes into account the inherent variability across collisions due to the presence of statistical fluctuations in the data.

With regard to the reconstruction pipelines of the experiments, we concentrate on a stage whereby individual particles have not yet been assigned to jets, i.e. to collections of final-state particles originating from the same scattered hard parton.

We expect the combined use of this technique with existing methods to result in further-improved performance in terms of pileup subtraction in higher-luminosity scenarios at the Large Hadron Collider, particularly with reference to the contamination from low-energy neutral particles. From a broader perspective, as more particle weighting methods are proposed, we envisage the possibility of combining the different weights, e.g. using multivariate techniques, with a view to making use of all the information available in the data with regard to which process individual particles originated from. It is also our opinion that the simplicity and parallelisation potential of this algorithm make it a promising candidate for inclusion in particle-level event filtering procedures upstream of jet reconstruction at future high-luminosity hadron collider experiments.

We intend to investigate possible ways of improving on the performance of this method, as well as to study more in detail the relation between this algorithm and the Markov Chain Monte Carlo technique that we used in a previous study where we proposed the idea of filtering individual collision events on a particle-by-particle basis at high-luminosity hadron colliders.

\section*{Acknowledgments}
The author wishes to thank the High Energy Physics Group at Brunel University London for a stimulating environment, and particularly Prof. Akram Khan, Prof. Peter Hobson and Dr. Paul Kyberd for fruitful conversations, as well as Dr. Ivan Reid for help with technical issues. Particular gratitude also goes to the High Energy Physics Group at University College London, especially to Prof. Jonathan Butterworth for his valuable comments. The author also wishes to thank Prof. Trevor Sweeting at the UCL Department of Statistical Science, as well as Dr. Alexandros Beskos at the same department for fruitful discussions. Finally, particular gratitude goes to Prof. Carsten Peterson and to Prof. Leif Lönnblad at the Department of Theoretical Physics, Lund University.


\end{multicols}

\begin{thebibliography}{99}
\bibitem{REVIEW_PILEUP_2014}The CMS Collaboration 2014 PAS JME-14-001
\bibitem{PUPPI}Bertolini D, Harris P, Low M and Tran N 2014 {\it J. High Energy Phys.} {\bf 1410}:059
\bibitem{SoftKiller}Cacciari M, Salam G~P and Soyez G 2014 ({Preprint} arXiv:1407.0408 [hep-ph]) 
\bibitem{berta}Berta P, Spousta M, Miller D~W and Leitner R 2014 ({Preprint} {\it J. High Energy Phys.} {\bf 1406}:092 
\bibitem{jet-sampling}Kahawala D, Krohn D and Schwartz M~D 2013 {\it J. High Energy Phys.} {\bf 1306}:006
\bibitem{gibbshep2}Colecchia F 2013 {\it J. Phys.: Conf. Ser.} {\bf 410} 012028
\bibitem{gibbshep}Colecchia F 2012 {\it J. Phys.: Conf. Ser.} {\bf 368} 012031
\bibitem{jet-area}Cacciari M and Salam G~P 2008 {\it Phys. Lett.} B {\bf 659}:119-26
\bibitem{jet-filtering}Butterworth J~M, Davison A~R, Rubin M and Salam G~P 2008 {\it Phys. Rev. Lett.} {\bf 100} 242001
\bibitem{jet-trimming}Krohn D, Thaler J and Wang L-T 2010 {\it J. High Energy Phys.} {\bf 1002}:084
\bibitem{jet-pruning-1}Ellis S~D, Vermilion C~K and Walsh J~R 2009 {\it Phys. Rev.} D {\bf 80}, 051501
\bibitem{jet-pruning-2}Ellis S~D, Vermilion C~K and Walsh J~R 2010 {\it Phys. Rev.} D {\bf 81}, 094023
\bibitem{jet-soft-drop}Larkoski A~J, Marzani S, Soyez G and Thaler J 2014 {\it J. High Energy Phys.} {\bf 1405}:146
\bibitem{jet-cleansing}Krohn~D, Schwartz~M~D, Low~M and Wang~L-T {\it Phys. Rev.} D {\bf 90}, 065020
\bibitem{pythia1}Sjöstrand T, Mrenna S and Skands P 2006 {\it J. High Energy Phys.} {\bf 0605}:026
\bibitem{pythia2}Sjöstrand T, Mrenna S and Skands P 2008 {\it Comput. Phys. Comm.} {\bf 178}
\bibitem{ETmiss}The CMS Collaboration 2013 {\it EPJC} {\bf 73} 2568
\end{thebibliography}
\end{document}